\documentclass[preprint,
 amsmath,amssymb,
 aps,nofootinbib
]{revtex4}

\usepackage{graphicx}
\usepackage{dcolumn}
\usepackage{bm}

\usepackage{float}
\usepackage{amsmath,amssymb}
\usepackage{graphicx}
\usepackage{color}

\usepackage{tikz}
\usetikzlibrary{hobby}
\usepackage{hyperref}

\usepackage{graphicx}
\usepackage{amsmath,amssymb,graphicx}
\usepackage{tabularx}
\newcolumntype{C}[1]{>{\centering\arraybackslash}p{#1}}
\usepackage{hyperref}
\usepackage{comment}
\usepackage{xcolor}
\usepackage{cleveref}
\newcommand{\be}{\begin{equation}}
\newcommand{\ee}{\end{equation}}
\newcommand{\bea}{\begin{eqnarray}}
\newcommand{\eea}{\end{eqnarray}}
\newcommand{\non}{\nonumber}

\begin{document}

\title{New bounds on vacuum decay in de Sitter space}

\author{Silvia Vicentini}

\email{silvia.vice06@gmail.com}
\affiliation{Dipartimento di Fisica, Universit\`{a} di Trento,}
\affiliation{Trento Institute for Fundamental Physics and Applications (TIFPA)-INFN,\\Via Sommarive 14, I-38123 Povo (TN), Italy}

\date{\today}

\begin{abstract}
Vacuum decay in de Sitter space is a process of great physical interest, as it allows to rule out cosmological models in the early and current Universe. Its rate may be described in terms of an instanton in Euclidean space called bounce and it is usually interpreted as thermally assisted quantum tunneling. According to analytical and numerical evidence in the literature, a bounce exists only for certain values of the Hubble parameter in single scalar field theories with Einstein-Hilbert gravity. In the present paper, we rely on a novel approach to provide more stringent bounds, which may be easily extended to theories with non-minimally coupled and quadratic gravity. An additional restriction is also derived, which specifically applies to the latter. 
\end{abstract}

\maketitle

\section*{Introduction}
Vacuum decay calculations trace back to the late '70s when Callan and Coleman found how to compute its rate using $O(4)$-symmetric instantons in spacetimes with Euclidean signature \cite{Coleman:1977py,Callan:1977pt}. It was found by qualitative arguments that they may be found as the trajectories bracketed between the so-called undershots and overshots \cite{Coleman:1977py}. Soon after that, Coleman and de Luccia considered gravitational effects on such phenomenon, which were computed for flat and de Sitter (dS) false vacua in the thin wall approximation \cite{Coleman:1980aw}. Now it is well established that in the latter case there is a rich zoology of spacetime-dependent instantons  \cite{Balek:2003uu,Jensen:1988zx,Jensen:1983ac,Battarra:2013rba,Weinberg:2006pc,Balek:2004sd,Hackworth:2004xb,Gregory:2020cvy}, which might be Coleman-de Luccia like (that is, monotonic functions of spacetime coordinates) or oscillating. This is related to two properties of de Sitter space. One is its compactness after a Wick rotation: instantons with many oscillations may be smoothly deformed into ones with fewer oscillations that still have finite action. The other corresponds to the effect of spacetime on scalar field trajectories, as it provides alternating patterns of undershots or overshots and, thus, possibly, many non-oscillating instantons. All these finite-action solutions in principle contribute to vacuum decay through bubble formation, despite this matter is still debated in the case of oscillating instantons \cite{Lavrelashvili:2006cv,Battarra:2012vu}). This means that one needs to compute and compare every decay rate to find which one dominates the process. As early as 1982, Hawking and Moss noted that no transition to the true vacuum should occur if the size of the resulting bubble (roughly the inverse of the scalar field mass) is larger than the cosmological horizon (the inverse of the Hubble parameter $H$), and proposed a competitive mechanism, the Hawking-Moss instanton \cite{Hawking:1981fz}. This qualitative bound was given careful consideration in the following years \cite{Balek:2003uu,Jensen:1983ac,Gregory:2020cvy,Balek:2004sd,Hackworth:2004xb}, as that would have excluded vacuum decay of scalar fields with even beyond-the-Standard-Model masses for inflationary values of the Hubble parameter. These results, which are both of theoretical and numerical nature, confirm the importance of the scalar field potential in setting bounds on $H$ such that quantum tunneling occurs. In particular, \cite{Balek:2003uu} investigated vacuum decay of scalar fields with Einstein-Hilbert gravity and found a sufficient and a necessary condition. The first one is 
 \bea
 \label{eq:chap2:suf}
\dfrac{3 V''(\phi_{\rm top}) M_P^2}{4 V(\phi_{\rm top})}+1<0
\eea 
where $V''(\phi_{\rm top})$ indicates the second scalar field derivative of the potential $V(\phi)$, computed at the top of the potential barrier separating the false vacuum from the true one (see Fig.\ref{fig:potential}, left panel). It corresponds to the condition such that undershots exist. While the authors infer that this is enough to guarantee the presence of a bounce, it was found numerically in \cite{Hackworth:2004xb} that overshots are not necessarily present. Thus, Eq.\eqref{eq:chap2:suf} guarantees the presence of instantons as long as they have at least one oscillation.\\
 The necessary condition instead is
\bea
\label{eq:chap2:nec}
\dfrac{V''(\phi)}{4 H^2}+1<0\qquad \text{with} \qquad H^2=\dfrac{V(\phi)}{3 M_P^2} \qquad \text{somewhere inside the barrier}
\eea
and it was furtherly supported by numerical evidence in \cite{Gregory:2020cvy,Balek:2004sd,Hackworth:2004xb}. Their calculation may be analogously carried out in scalar field theories on a fixed de Sitter background, giving again Eq.\eqref{eq:chap2:nec} in which though $H$ is a free parameter. In this case, Eq.\eqref{eq:chap2:nec} places an upper bound on $H$ such that instantons exist.\\
 It has also been found numerically \cite{Jensen:1983ac} that, if
 \bea
 \label{eq:chap2:neqtop}
\left| V''(\phi_{\rm top})\right|\neq 0,
 \eea
 and $V''(\phi)$ is monotonically decreasing for $\phi_{\rm fv}<\phi<\phi_{\rm top}$, the Coleman-de Luccia bounce disappears for values of $H$ such that 
  \bea
\label{eq:leftbound}
 \dfrac{V''(\phi_{\rm top})}{4 H^2}+1
 \eea
 vanishes.
 This, along with Eq.\eqref{eq:chap2:nec}, lead the authors to think that Eq.\eqref{eq:leftbound} should be negative. Nonetheless, it was numerically shown that, if the potential has a negative quartic derivative in $\phi_{\rm top}$, the Coleman-de Luccia bounce exists for positive values of Eq.\eqref{eq:leftbound} \cite{Hackworth:2004xb} and not negative ones. In particular, in both cases there are indeed undershoot trajectories, but overshoot ones only appear in the latter. These findings thus suggest that, if Eq.\eqref{eq:chap2:neqtop} holds, and $V''(\phi)$ is monotonically decreasing for $\phi_{\rm fv}<\phi<\phi_{\rm top}$, the Coleman-de Luccia instanton do not exists for 
 \begin{equation}
  \begin{cases}
  \label{eq:num}
V''(\phi_{\rm top})+4H^2>0&\text{for}\qquad V''''(\phi_{\rm top})>0\\
 V''(\phi_{\rm top})+4H^2<0&\text{for}\qquad V''''(\phi_{\rm top})<0.\\
 \end{cases}
 \end{equation}
 \vspace{0.5cm}
 
 There are a number of open issues that are worth addressing. First of all, part of such numerical evidence (such as Eq.\eqref{eq:num})  is not yet supported theoretically. Also, Eq.\eqref{eq:chap2:nec} does not apply to Higgs decay, as there is no true vacuum and thus it is not clear how far the potential barrier extends. Moreover, it gets infinite and negative at $V(\phi)=0$ and thus no bound arises. In addition, it would be interesting to determine how this bound shifts when a non-minimal coupling to the Ricci scalar $R$ is included in the theory. For example, $\xi \phi^2 R$ behaves as a scalar field mass term for constant $R$ (that is de Sitter) values, thus possibly contributing to $V''(\phi)$ in Eq.\eqref{eq:chap2:nec}. Nonetheless, to our knowledge, no generalization of Eq.\eqref{eq:chap2:nec} to theories non-minimally coupled to gravity has been put forward so far. Moreover, quadratic gravity terms $\alpha R^2$ have an additional (gravitational) scalar degree of freedom, which might impose similar restrictions to the scalar field ones. Again, this scenario has not been explored yet. \\
 \vspace{0.5cm}
 
 In this paper, we present a novel approach that allows us to address these issues. As the contribution of oscillating instantons to vacuum decay is uncertain, we focus on non-oscillating ones. For brevity, we will refer to them as Coleman-de Luccia instantons or bounces. We consider first scalar field decay on a fixed de Sitter background (Sect.\ref{sec:backgrounddS}). In this way, we are able to recover Eq.\eqref{eq:chap2:nec}, which sets an upper bound on $H$ such that bounces exist. We also find a lower bound on $H$ which partially overlaps with the numerical evidence presented above. Our results are applied to a toy model and to Higgs decay, for which a lower bound on $H$ is found: this allows us to rule out Higgs decay for current values of the cosmological constant. This calculation is extended to theories with Einstein-Hilbert gravity in the case of small gravitational backreaction (Sect.\ref{sec:EHds}), which is expressed in terms of scalar field scales being smaller than the Planck mass. Then we consider generalizations of such calculations to theories with a non-minimal coupling $\xi \phi^2 R$ and a quadratic Ricci scalar $\alpha R^2$. We focussed on such terms because they are required by perturbative renormalizability in quantized field theories on a gravitational (classical) background \cite{birrell,parker,Buchbinder:1992rb}. Moreover, the non-minimal coupling is crucial in Higgs inflation (see \cite{Rubio:2018ogq,Wetterich:2019qzx} and references therein) while the quadratic term allows solving the strong coupling problem that affects it  \cite{Ema:2020evi,Ema:2017rqn,Gorbunov:2018llf}. They appear also in Starobinsky \cite{Starobinsky:1980te} and scale-invariant inflation \cite{GarciaBellido:2011de,Karananas:2016kyt,Ferreira:2018qss,Rinaldi:2015uvu}. We find that our calculation straightforwardly extends to theories with a non-minimal coupling (Sect.\ref{sec:nonm}) and that no Coleman de Luccia bounce exists for $\xi \geq 1/3$ which is also confirmed by numerical calculations in two toy models. We also find that Eq.\eqref{eq:chap2:nec} readily applies also when a quadratic gravity term is included. As it will be clear in the following, this method may be equivalently applied to the gravitational scalar degree of freedom also, and then analogous bounds apply. Anyway, calculations are cumbersome as its equation of motion is coupled to matter by the stress-energy tensor. For this reason, we take a step back and consider instead the properties of scalar degrees of freedom in de Sitter vacuum decay, in order to apply them to the gravitational one. This allows to derive a new bound, which is discussed in Sec.s\ref{sec:chap4:quadr}-\ref{sec:chap4:quadrnonm}. Conclusions follow.
\section{A new criterion for vacuum decay in de Sitter space}
 \label{sec:backgrounddS}
 \subsection{Instantons in de Sitter space}
 \label{sec:bdsintro}
\begin{figure*}
\centering
\mbox{
\begin{tikzpicture} [scale=2.8]
\draw (0,0.1) to [curve through={(0.3,0)..( 0.6,0.3)..(0.8,0.7)..(1.02,0.5)..(1.12,0)..(1.14,-0.1)..(1.3,-0.35)..(1.4,-0.38)..(1.6,-0.34)..(1.8,-0.17)}](2.1,0.3);
\draw[-stealth] (0,-1.) to (0,1.3);
\draw[-stealth] (0,-0.2) to (2.5,-0.2);
\node at (0.2,1.2) [scale=1] {$V(\phi)$};
\node at (2.45,-0.3) [scale=1] {$\phi$};
\node at (0.3,-0.1) [scale=1] {$\phi_{\rm fv}$};
\node at (1.43,-0.52) [scale=1] {$\phi_{\rm tv}$};
\end{tikzpicture}
}
 \caption{Scalar field potential with two classical vacua.  $\phi_{\rm fv}$ is the value of the scalar field at the false vacuum, separated from the local minimum of the potential by a barrier, on top of which $\phi=\phi_{\rm top}$.} 
    \label{fig:potential}
\end{figure*}
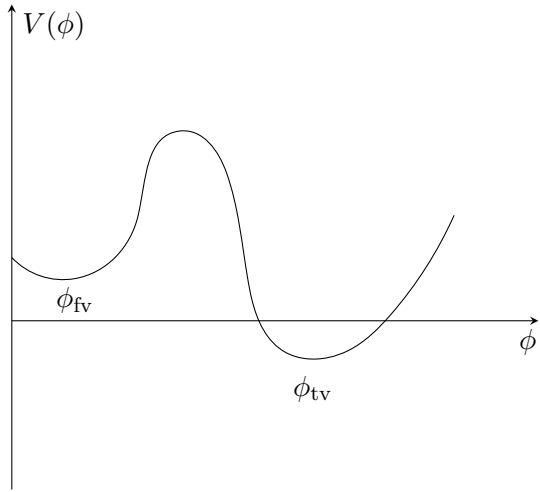

 The decay of metastable states due to quantum tunnelling occurs at an exponentially small rate, given by 
\bea
\Gamma=A e^{-B}
\eea
$A$ and $B$ may be computed in Euclidean space as described in  \cite{Coleman:1977py,Callan:1977pt}. The latter in particular is called the tunneling exponent and it controls strong enhancements and suppressions of $\Gamma$. It corresponds to the difference between the Euclidean action of the theory calculated on an instanton solution and the one computed in the false vacuum state. Consider an $O(4)$-symmetric spacetime, described by line element
\bea
\label{eq:line}
ds^2=dt^2+\rho(t)^2d\Omega_3^2
\eea
and an $O(4)-$symmetric single scalar field theory with a metastable state (as $\phi_{\rm fv}$ in Fig.\ref{fig:potential}) on such spacetime with action
\bea
\label{eq:action}S=2 \pi^2\int_{0}^{+\infty} dt\,\rho(t)^3\left(\frac{\dot\phi^2}{2}+V(\phi)\right)\,,
\eea
where the dot indicates a derivative with respect to the Euclidean time $t$, and $\phi=\phi(t)$. The areal radius $\rho(t)$ is customarily called the scale factor and it corresponds to 
\bea
\label{eq:scaleds}
\rho(t)=\dfrac{\sin(H t)}{H}
\eea
in de Sitter space, where $H$ is the Hubble parameter and it is a free parameter of the theory. Spacetime extends from $t=0$ to $t=\pi H^{-1}$ and thus it is compact. Instantons (i.e. trajectories which keep the on-shell action finite) are solution to the equation of motion
 \bea
 \label{eq:eom}
 \ddot{\phi}+\dfrac{3 H \cos(Ht)\dot{\phi}}{\sin(Ht)}=\dfrac{dV}{d\phi}
 \eea 
 with boundary conditions
 \bea
 \label{eq:bc}
\lim_{t\rightarrow\pi H^{-1}} \dot \phi(t)=0\qquad \dot\phi(0)=0.
\eea
To make the velocity vanish at finite values of $t$, the instanton should have finite acceleration there. If the potential barrier extends to the right of the false vacuum state (as in Fig.\ref{fig:potential}), Coleman-de Luccia instantons have negative velocity and thus (using Eq.\eqref{eq:eom})
\bea
\ddot\phi(\pi H^{-1})\equiv\dfrac{dV}{d\phi}(\bar{\phi})>0\qquad \text{with}\qquad \bar{\phi}\equiv\phi(\pi H^{-1}).
\eea
Notice also that the second term on the left-hand side of Eq.\eqref{eq:eom} provides friction for $Ht\in [0,\pi/2]$ and self-acceleration for $Ht\in [\pi/2, \pi H^{-1}]$. So, there is possibly an alternating pattern of undershots and overshots as a function of the initial conditions $\phi(0)$ and, thus, multiple instantons. The one with the smallest tunneling exponent dominates decay, which means that it expresses the preferred escaping path of particles through the potential barrier via quantum tunneling \cite{Markkanen:2018pdo}. Thus, if instead no instanton is found, the tunneling process does not take place.\\

In the following, we consider scalar field theories as in Eq.\eqref{eq:action} on a fixed de Sitter spacetime, in order to derive bounds on $H$ such that no bounce exist. Notice that the methods adopted in \cite{nostro1,nostro2} for vacuum decay in flat space cannot be used in this case. In fact, they consist of a spacetime boundary analysis (near $t\rightarrow +\infty$) which relies on it being non-compact. We consider instead the following change of variables
\bea
\label{eq:chap3:fpi2}
\psi(Ht)=\dfrac{\phi(Ht)-\phi(\pi/2)}{\cos(H t)}
\eea
that relates an instanton solution $\phi(t)$ to another one $\psi(t)$. Taking derivatives in $t$ and setting $H t=\pi/2$ one finds 
\begin{gather}
\label{eq:dotfpi2}
\dot{\psi}(\pi/2)=-\dfrac{\ddot{\phi}(\pi/2)}{2H}\qquad \ddot\psi(\pi/2)=-\dfrac{\dot\phi(\pi/2)}{3 H}\left(4 H^2+\dfrac{d^2V}{d\phi^2}\right).
\end{gather}
Then, bounds like Eq.\eqref{eq:chap2:nec} might readily emerge from an analysis of the $\psi$ dynamics.
 However, a $\phi$ bounce may correspond to an oscillating $\psi$ instanton. Nonetheless, the number of turning points for $\psi$ ($t^*$ such that $\dot{\psi}(t^*)=0$) is preserved on the bounce as $H$ changes, as, otherwise, infinite energy (i.e., non-instantonic) configurations would be crossed.\\ We now show (Sect.\ref{sec:monotonic}-\ref{sec:oscillating}) that, if
 \bea
 \label{eq:leftboundbarrier}
 V''+4 H^2 \phi
 \eea
 is of definite sign throughout the potential barrier, that is if
 \bea
 \label{eq:chap3:vds}
 V'+4 H^2 (\phi-\phi_{top})
 \eea
 is monotonic, then there are no Coleman-de Luccia $\phi-$instantons. To do that, we show that no $\psi$ instantons exist. As the number of turning points of $\psi$ on the bounce is undetermined, we need to consider both oscillating and Coleman-de Luccia $\psi$ instantons. Let us start with the Coleman-de Luccia ones.
 \subsection{Coleman-de Luccia $\psi-$instantons}
 \label{sec:monotonic}
  One can easily see that, if Eq.\eqref{eq:leftboundbarrier} is of definite sign throughout the potential barrier, there are no Coleman-de Luccia $\psi$-instantons. $\ddot{\psi}(0)$ and $\ddot{\psi}(\pi)$ read
\bea
\label{eq:ddotpsi}
\ddot{\psi}(0)=V'(\phi(0))+4H^2(\phi(0)-\phi(\pi/2)),\\
\ddot{\psi}(\pi)=-V'(\phi(\pi))-4H^2(\phi(\pi)-\phi(\pi/2)),
\eea 
and thus
\bea
\begin{cases}
\ddot{\psi}(\pi)>0\qquad \text{for}\qquad \phi(\pi/2)>\phi_{\rm top}\\
\ddot{\psi}(0)>0\qquad \text{for} \qquad \phi(\pi/2)<\phi_{\rm top}
\end{cases}
\eea
if Eq.\eqref{eq:chap3:vds} is monotonically increasing (see the top panel in Fig.\ref{fig:examplepoints}). If there are no turning points, this gives
\bea
\label{eq:checkpi2}
\begin{cases}
\dot{\psi}(Ht)>0\qquad \text{for} \qquad\phi(\pi/2)>\phi_{\rm top}\\
\dot{\psi}(Ht)<0\qquad \text{for} \qquad\phi(\pi/2)<\phi_{\rm top}.
\end{cases}
\eea
for all $t$. However, $\dot{\psi}(\pi/2)$ is given by Eq.\eqref{eq:dotfpi2} contradicting Eq.\eqref{eq:checkpi2}. Analogously, one can prove that there are no Coleman-de Luccia $\psi-$instantons with negative  Eq.\eqref{eq:leftboundbarrier}.
 \subsection{Oscillating $\psi-$instantons}
 \label{sec:oscillating}
\begin{figure}
     
       \mbox{
      \begin{minipage}{0.44\textwidth}
     \includegraphics[scale=0.6]{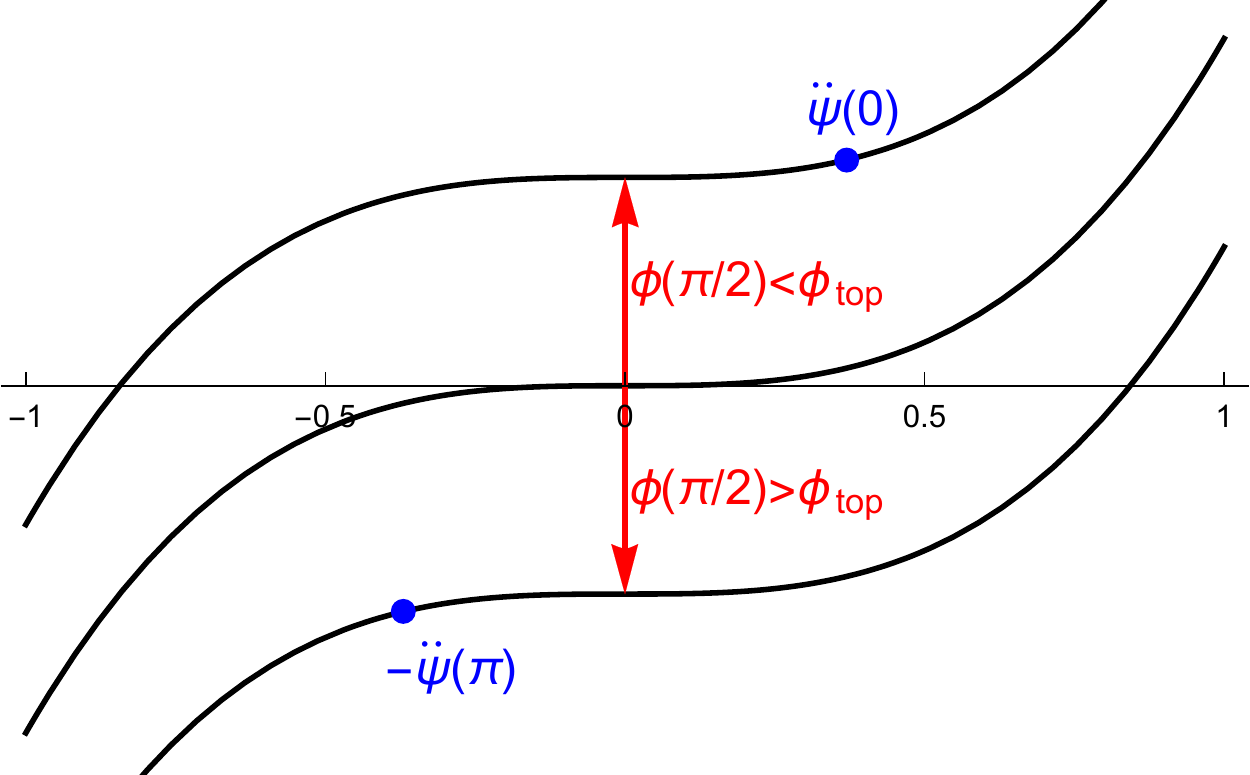}
     \end{minipage}
       \begin{minipage}{0.26\textwidth}
     \includegraphics[scale=0.3]{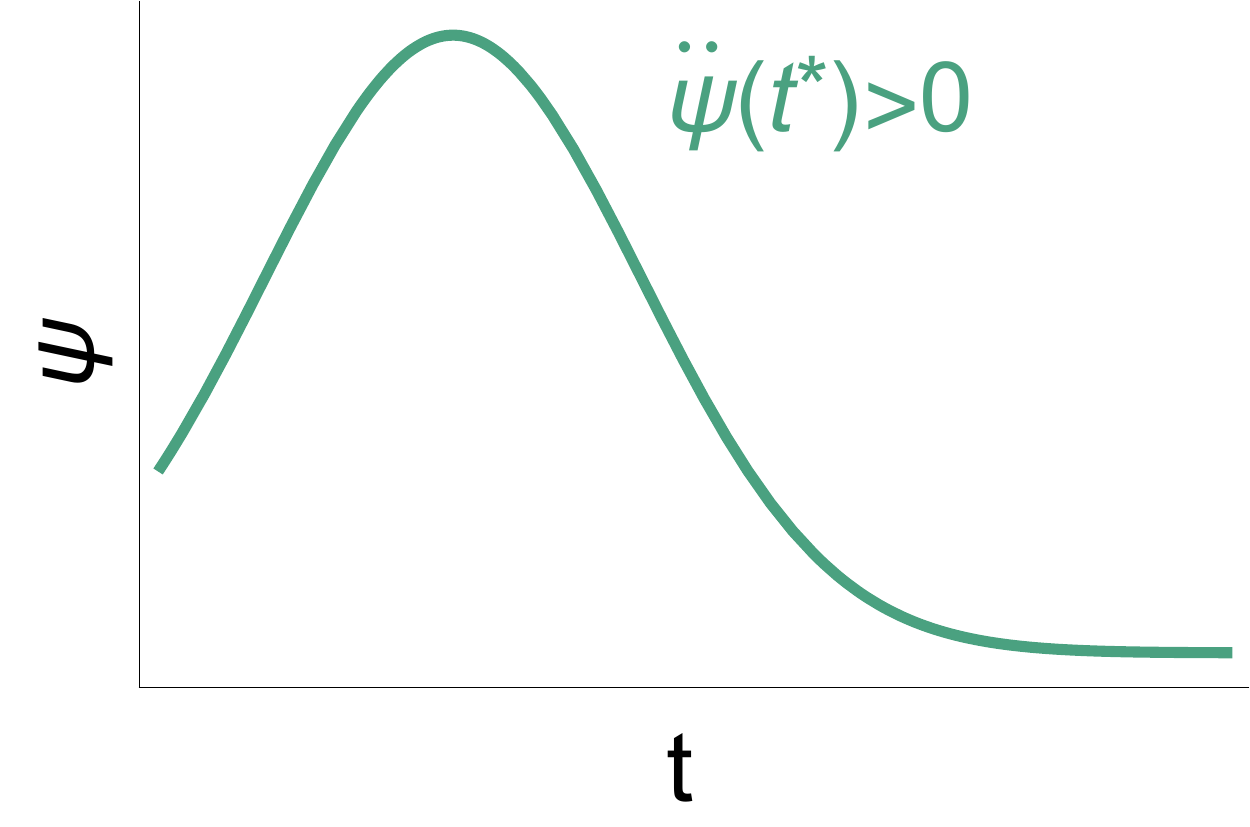}
     \end{minipage}
     \begin{minipage}{0.3\textwidth}
     \includegraphics[scale=0.3]{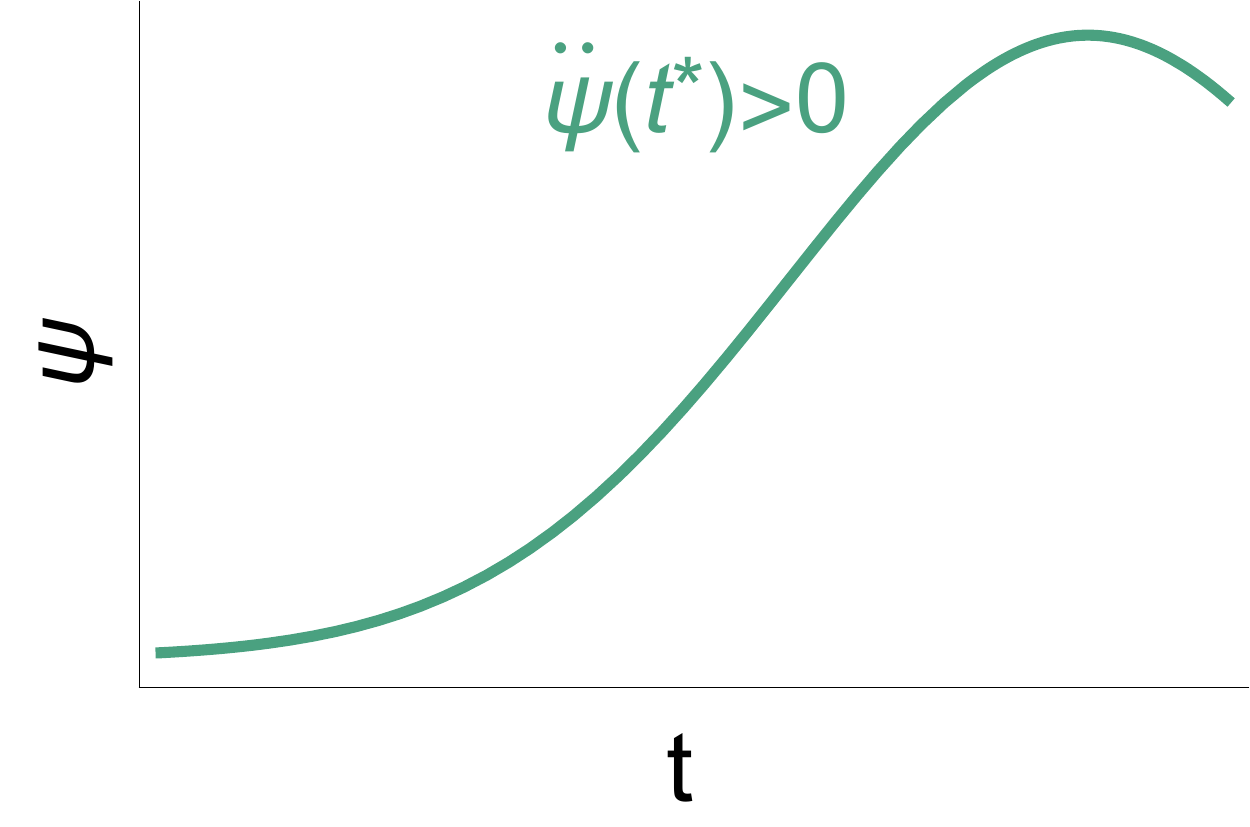}
     \end{minipage}}
    \mbox{
      \begin{minipage}{0.44\textwidth}
     \includegraphics[scale=0.6]{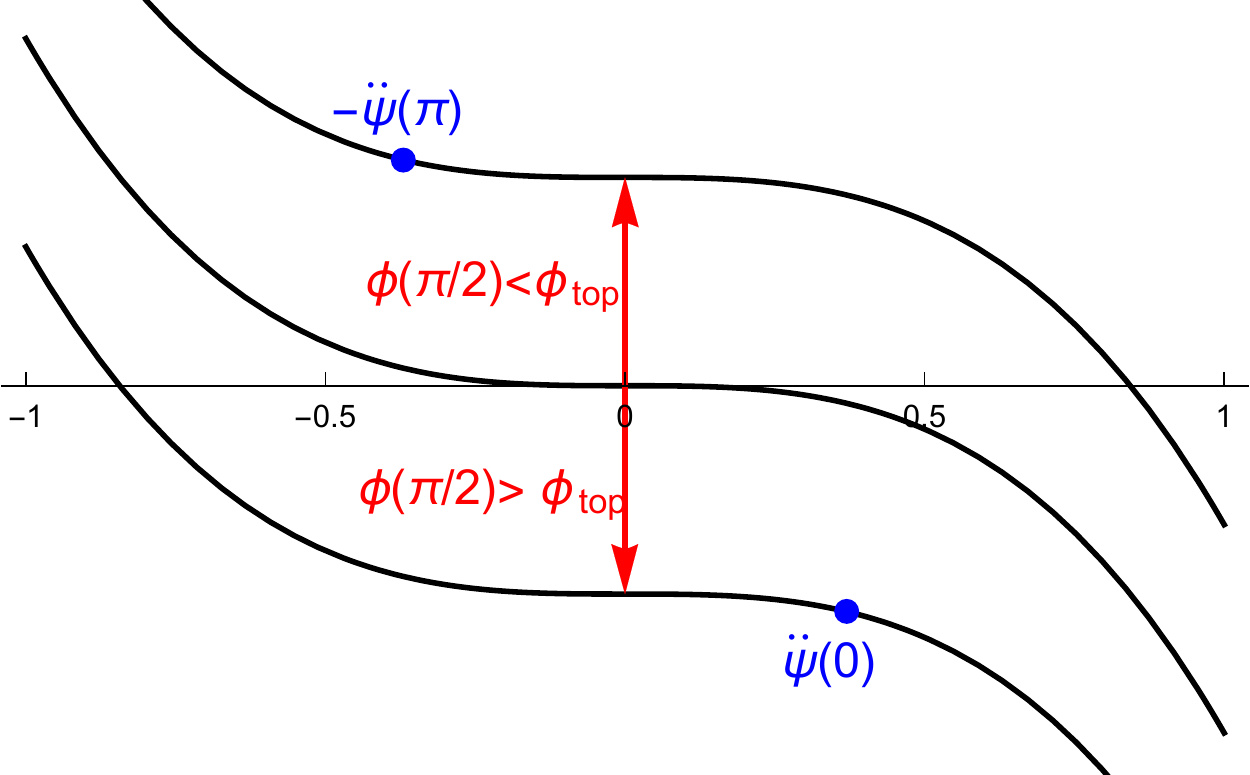}
     \end{minipage}
       \begin{minipage}{0.26\textwidth}
     \includegraphics[scale=0.3]{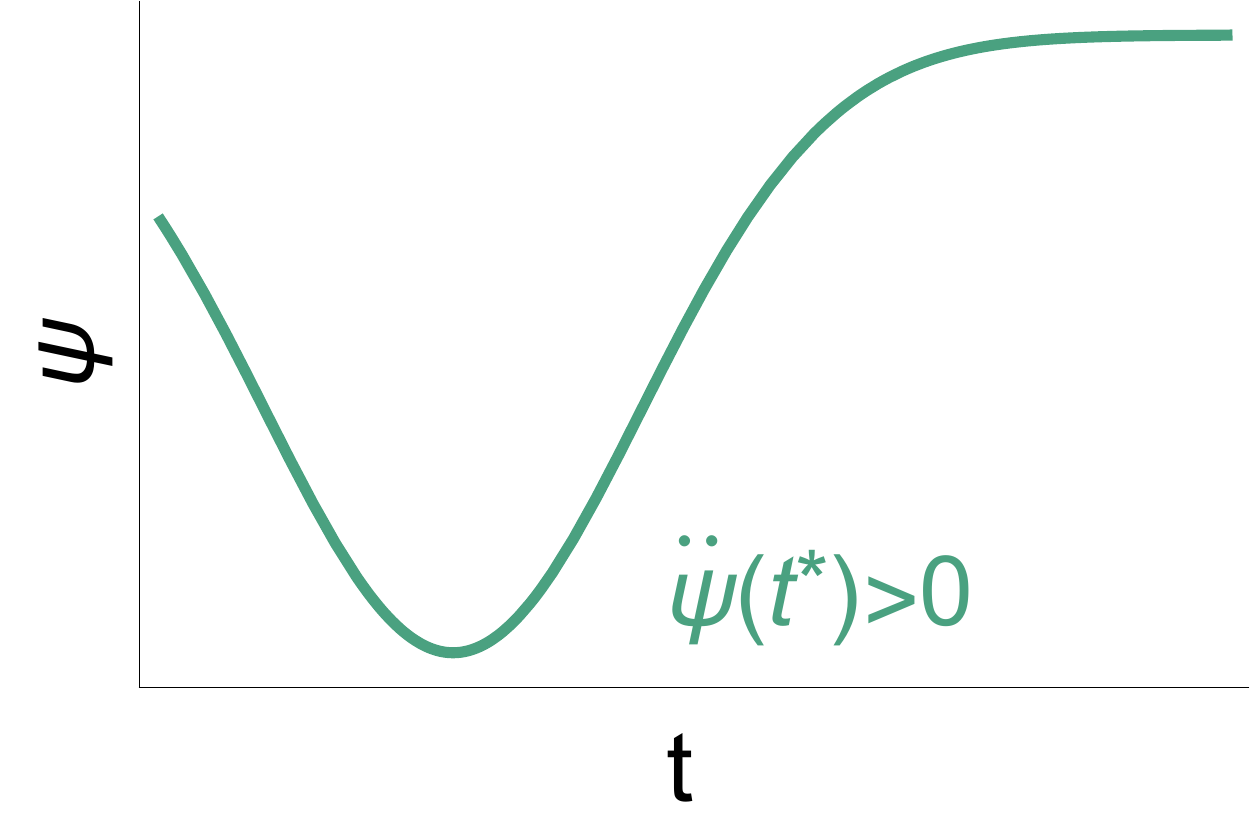}
     \end{minipage}
     \begin{minipage}{0.3\textwidth}
     \includegraphics[scale=0.3]{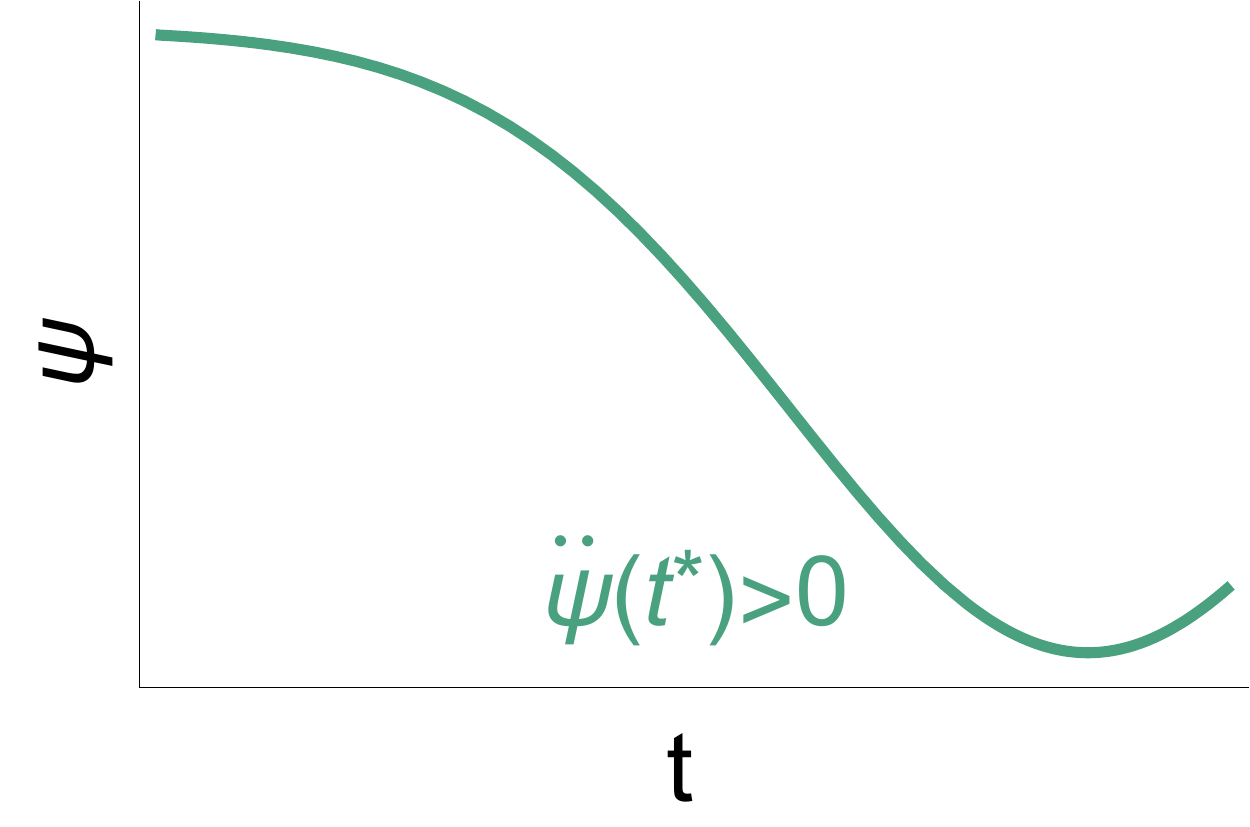}
     \end{minipage}}
     \caption{Eq.\eqref{eq:chap3:vds} as a function of $\phi$ (in black) with initial/final points for $\psi-$instantons, with monotonically increasing (upper panel ,left) and monotonically decreasing (bottom panel,left) Eq.\eqref{eq:chap3:vds}. On the right one has the first (centre) and last (right) turning point in each case. Conditions on the instanton velocity and acceleration are reported in green.}
     \label{fig:examplepoints}
 \end{figure}

 These instantons have turning points, and thus there is at least one value of $t$, say $t^*$, such that $\dot\psi (H t^*)=0.$ The acceleration at this point is given by
\bea
\label{eq:chap3:turnf}
\ddot{\psi}(Ht^*)=\dfrac{1}{\cos(Ht^*)} \left(V'(\phi)+4H^2(\phi(Ht^*)-\phi(\pi/2))\right).\eea
If negative, we have that 
\begin{equation}
\begin{cases}
\label{eq:condds1}
V'(\phi(Ht^*))+4H^2\phi(Ht^*)<4H^2\phi(\pi/2)&\text{for}\qquad 
Ht^*<\pi/2\\ V'(\phi(Ht^*))+4H^2\phi(Ht^*)>4H^2\phi(\pi/2)& \text{for}  \qquad 
Ht^*>\pi/2\\
V''(\phi(\pi/2))+4 H^2<0&  \text{for}  \qquad 
Ht^*=\pi/2.
\end{cases}
\end{equation}
The latter in particular requires $\dot{\psi}(\pi/2)=0$, i.e. $\phi(\pi/2)=\phi(H t^*)=\phi_{\rm top}$ according to Eq.\eqref{eq:dotfpi2}.
If positive 
\begin{equation}
\begin{cases}
\label{eq:condds2}
V'(\phi(Ht^*))+4H^2\phi(Ht^*)>4H^2\phi(\pi/2)&\text{for}\qquad 
Ht^*<\pi/2\\ V'(\phi(Ht^*))+4H^2\phi(Ht^*)<4H^2\phi(\pi/2)& \text{for}  \qquad 
Ht^*>\pi/2\\
V''(\phi(\pi/2))+4 H^2>0&  \text{for}  \qquad 
Ht^*=\pi/2.
\end{cases}
\end{equation}
Consider the case in which 
\bea
\label{eq:condnec}
V''(\phi)+4 H^2>0
\eea
throughout the potential barrier, then
\bea
\begin{cases}
\label{eq:final1}
V'(\phi)+4H^2\phi\geq4H^2\phi_{\rm top}\qquad\text{for}\qquad V'(\phi)\leq0\\
V'(\phi)+4H^2\phi<4H^2\phi_{\rm top}\qquad\text{for}\qquad V'(\phi)>0.\\
\end{cases}
\eea
If $\dot{\psi}(\pi/2)<0$ on the bounce, then $\phi(\pi/2)<\phi_{\rm top}$ according to Eq.\eqref{eq:dotfpi2}, which gives $\ddot\psi (0)>0$  (see Fig.\ref{fig:examplepoints}, top panel). Then, the first turning point has $\ddot\psi(H t^*)<0$ and $\dot \psi (t)>0$ for $t\in [0,t^*]$, so necessarily $Ht^*<\pi/2$ and
\bea
\phi(H t^*)>\phi(\pi/2)
\eea
giving 
\bea
\label{eq:dVtstar}
V'(\phi(Ht^*))<0
\eea
according to Eq.\eqref{eq:condds1}. Moreover,
\bea
\label{eq:slopedV12}
V'(\phi(Ht^*))+4H^2\phi(Ht^*)<4H^2\phi_{\rm top}.
\eea
But Eq.\eqref{eq:slopedV12} and Eq.\eqref{eq:dVtstar} contraddict Eq.\eqref{eq:final1}, and than no instantons with turning points exist if Eq.\eqref{eq:leftboundbarrier} is positive throughout the potential barrier and $\dot\psi(\pi/2)$ is negative. Analogously, if $\dot\psi(\pi/2)>0,$ then $\ddot{\psi(\pi)}>0$  (Fig.\ref{fig:examplepoints}, top panel) and the last turning point has $\ddot{\psi}(H t^*)<0$. By the same arguments, we find again
\bea
V'(H t^*)<0\qquad 
V'(\phi(Ht^*))+4H^2\phi(Ht^*)<4H^2\phi_{\rm top}
\eea
in disagreement with Eq.\eqref{eq:final1}.\\
In the same way, if
\bea
\label{eq:condlow}
V''(\phi)+4H^2<0
\eea
throughout the potential barrier, then $\ddot{\psi}(0)<0$ for positive $\dot\psi(\pi/2)$, or $\ddot{\psi}(\pi)<0$ for negative $\dot\psi(\pi/2)$, contradicting Eq.\eqref{eq:condlow} (Fig.\ref{fig:examplepoints}, bottom panel) .
\vspace{0.5cm}

In summary, if Eq.\eqref{eq:leftboundbarrier} has definite sign throughout the potential barrier, there are no Coleman-de Luccia instantons. Then, Coleman-de Luccia instantons do not exist for
\bea
\begin{cases}
H\geq H_{\rm b}&\text{with}\qquad V''(\phi)+4 H_{\rm b}^2\geq 0 \qquad \text{in the whole potential barrier}\\
H\leq H_{\rm b}&\text{with}\qquad V''(\phi)+4 H_{\rm b}^2\leq 0 \qquad \text{in the whole potential barrier}.
\end{cases}
\eea
This improves previously existing results which state that there are no instantons if Eq.\eqref{eq:leftboundbarrier} is positive definite. As a final comment, notice that the proof holds in the more general setting requiring only Eq.\eqref{eq:chap3:vds} to be of definite sign on either sides of the potential barrier. This fact will be used in Sect.\ref{sec:nonm} to extend our findings to non-minimally coupled theories.\\
\subsection{Bounds at $\phi_{\rm top}$}
\begin{figure}
\centering
     \includegraphics[scale=0.6]{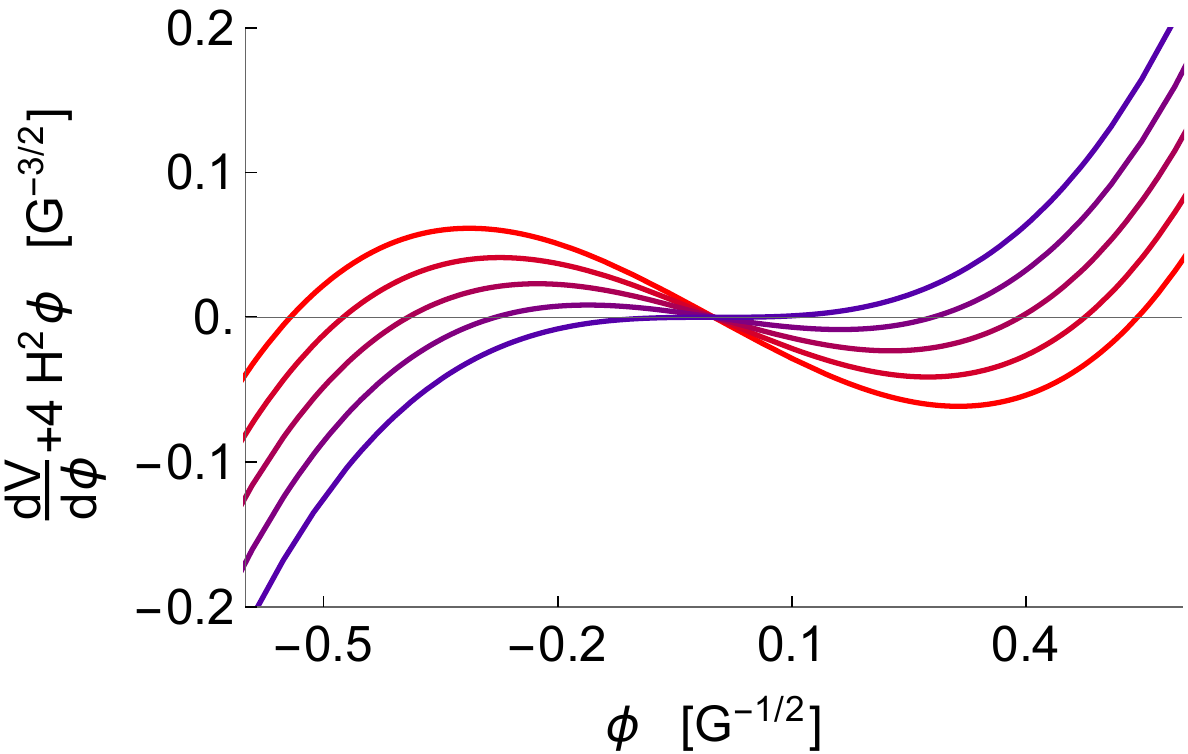}
    
     \caption{Eq.\eqref{eq:chap3:vds} with $V(\phi)=-0.5 \phi^2+0.25 \phi^4$ and $H$ from $H=0.4$ (red) to $H=H_{\rm b}=0.5$ (blue). For $H\leq H_{\rm b}$ a small region around $\phi_{\rm top}=0$ opens up allowing for $\psi-$instantons.}
     \label{fig:boundH}
 \end{figure}

In this section we bridge the results outlined above with numerical findings reported in the literature and summarized in Eq.\eqref{eq:num} in the Introduction. In fact, such bound looks similar, apart from the fact that Eq.\eqref{eq:leftboundbarrier} is computed at $\phi=\phi_{\rm top}$. Start by considering a more restricting condition than the one in the previous section, taking
\bea
\begin{cases}
\label{eq:boundtop}
V''(\phi)+4 H^2_{\rm b}=0&\text{for}\qquad \phi=\phi_{\rm top}\\
V''(\phi)+4 H^2_{\rm b}\neq 0&\text{for}\qquad \phi\neq \phi_{\rm top}
\end{cases}
\eea
This means that, for $H$ close to $H_{\rm b}$, the non-monotonicity of Eq.\eqref{eq:chap3:vds} is confined to a region $\phi\sim \phi_{\rm top}$ (see Fig.\ref{fig:boundH}).
 Then, as $H$ is increased towards $H_{\rm b}$,
 \begin{itemize}
 \item $\psi$-instantons with no turning points have initial conditions in the region breaking monotonicity, and thus $\phi(0)$ is pushed towards $\phi_{\rm top}$ as $H\rightarrow H_{\rm b}$;
 \item $\psi-$instantons with turning points have $\psi(H t^*)$ pushed towards $\phi_{\rm top}$ and, by regularity of Eq.\eqref{eq:chap3:turnf}, also $\phi(\pi/2)$ does. Then 
 \bea
  \dot{\psi}(\pi/2)=0\qquad \ddot{\psi}(\pi/2)=0
 \eea
 for $H=H_{\rm b}$.
 \end{itemize}
 Thus, in the limit that Eq.\eqref{eq:leftbound} holds, one finds that the only solution with a turning point is the static one $\phi(t)=\phi_{\rm top}$.
 \vspace{0.5cm}

\begin{figure}
  \centering
     \mbox{
     \begin{minipage}{0.5\textwidth}
     \includegraphics[scale=0.6]{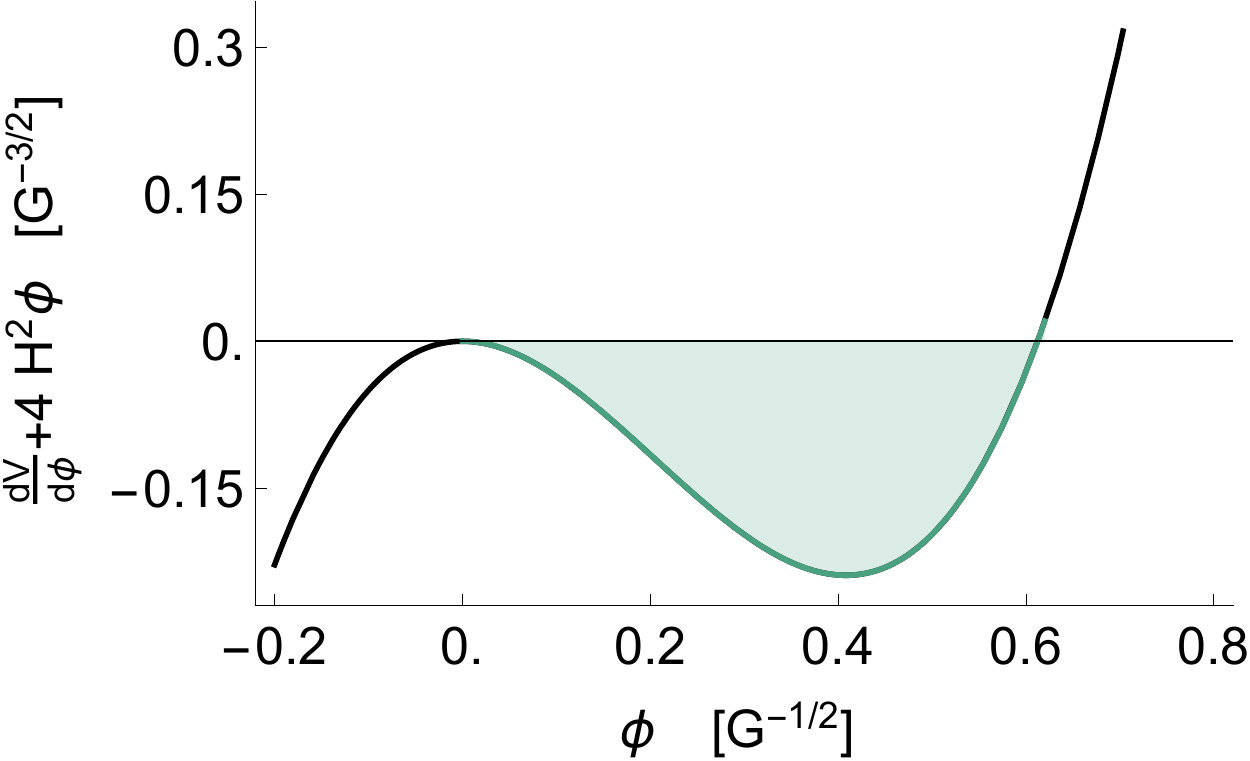}
     \end{minipage}
     \begin{minipage}{0.5\textwidth}
     \includegraphics[scale=0.6]{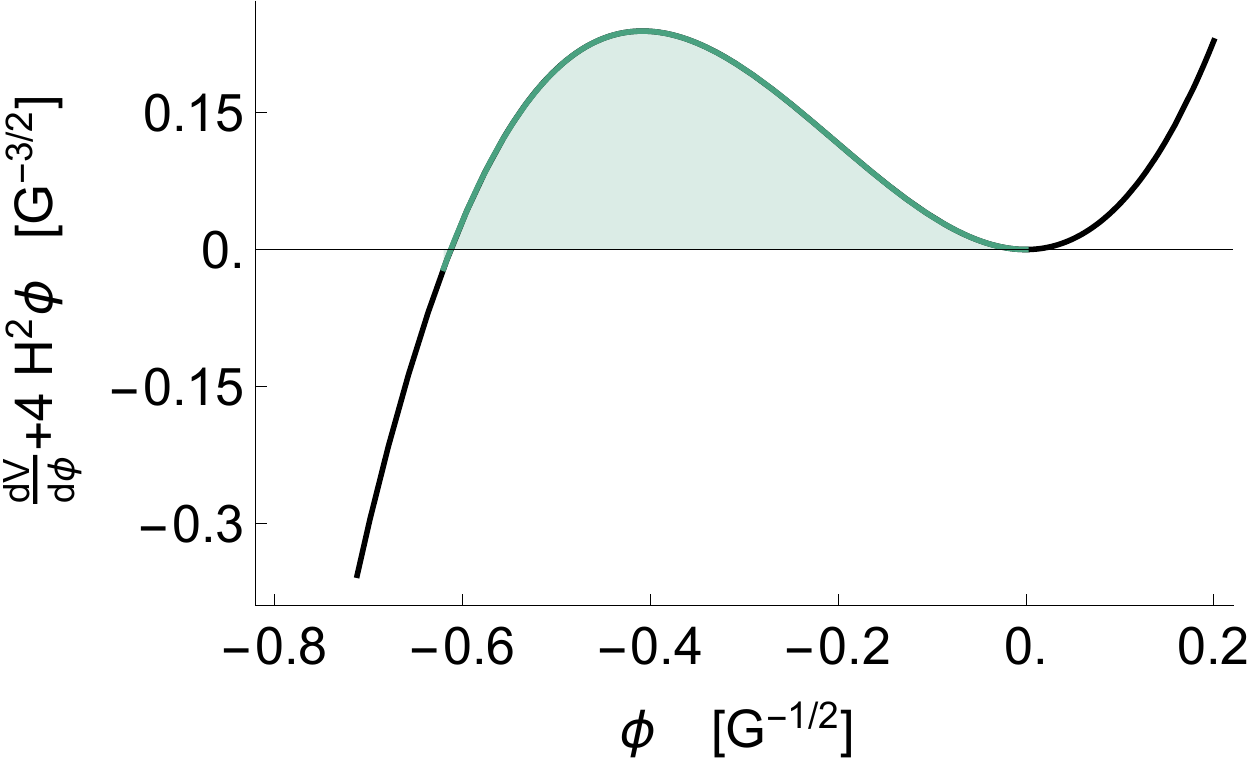}
     \end{minipage}}
     \caption{Eq.\eqref{eq:chap3:vds} for a polynomial potential Eq.\eqref{eq:quadrpotential} with $b=-\frac{1}{2\sqrt{6}}$ on the left, $b=\frac{1}{2\sqrt{6}}$ on the right and $c=1.$ The green region highlights the range of scalar field values for which the monotonicity condition breaks, and it lies at $\phi<\phi_{\rm top}$ ($\phi>\phi_{\rm top}$) for $b<0$ ($b>0$)}
     \label{fig:cubicds}
     \end{figure}

Actually, hitting a stationary point may occur also if Eq.\eqref{eq:chap3:vds} is not monotonic. Consider the case in which $H=H_{\rm b}$ (which is determined by the first of Eq.\eqref{eq:boundtop}), but monotonicity is broken around $\phi_{\rm top}$ (see Fig.\ref{fig:cubicds}). If the monotonicity condition is softly broken\footnote{This may be measured by comparing the range of scalar field values breaking the monotonicity condition with the width of the potential barrier $[\phi_{\rm fv},\phi_{\rm tv}]$}, then the bounce trajectory is close to the monotonicity-preserving one, and then $\phi(\pi/2)\rightarrow\phi_{\rm top}$. Then, from Eq.\eqref{eq:dotfpi2} one has 
\bea
  \dot{\psi}(\pi/2)=0\qquad \ddot{\psi}(\pi/2)=0.
 \eea
If the monotonicity condition breaks for $\phi>\phi_{\rm top}$ (see for example Fig.\ref{fig:cubicds}, on the left), then instantons have $\dot{\psi}(\pi/2)<0$ and $\ddot \psi(\pi)>0$, i.e. $\phi(\pi/2)<\phi_{\rm top}$. If instead it breaks for $\phi<\phi_{\rm top}$ (Fig.\ref{fig:cubicds}, on the right), one has instantons with $\phi(\pi/2)>\phi_{\rm top}$. If the quartic derivative of $V(\phi)$ in $\phi$ is positive (negative) definite in $[\phi_{\rm fv},\phi_{\rm tv}]$, then the monotonicity requirement is broken only on the right or on the left of $\phi_{top}$ (depending on other terms in the potential) for $H=H_{\rm b}$, as Eq.\eqref{eq:leftboundbarrier} has definite curvature. If such breaking is additionally small, there is no bounce for 
\begin{equation}
  \begin{cases}
V''(\phi_{\rm top})+4H^2>0&\text{for}\qquad V''''(\phi_{\rm top})>0\\
 V''(\phi_{\rm top})+4H^2<0&\text{for}\qquad V''''(\phi_{\rm top})<0.\\
 \end{cases}
 \end{equation} and one recovers the numerical evidence reported in the Introduction.
\subsection{Examples}

\begin{figure}
\centering\mbox{\begin{minipage}{0.5\textwidth}
\quad\, \includegraphics[scale=0.8]{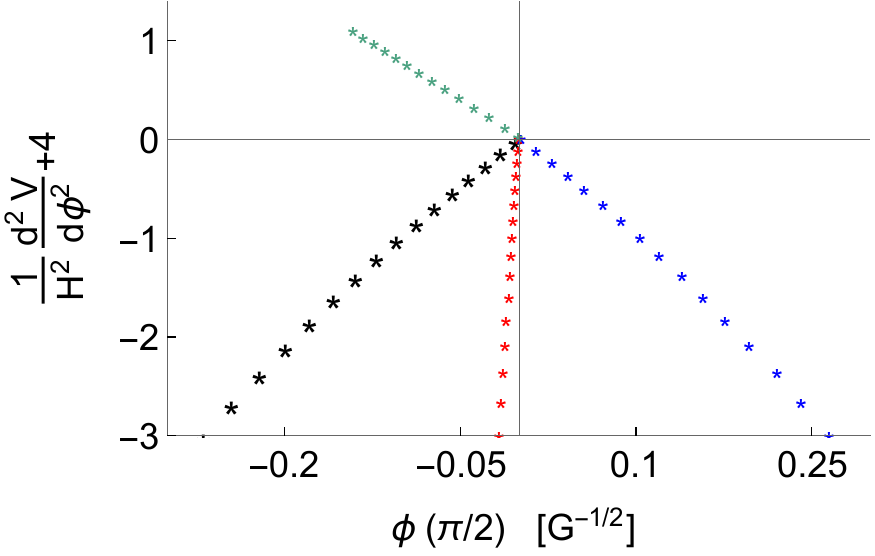}\end{minipage}\quad\,\,\begin{minipage}{0.5\textwidth}
 \includegraphics[scale=0.8]{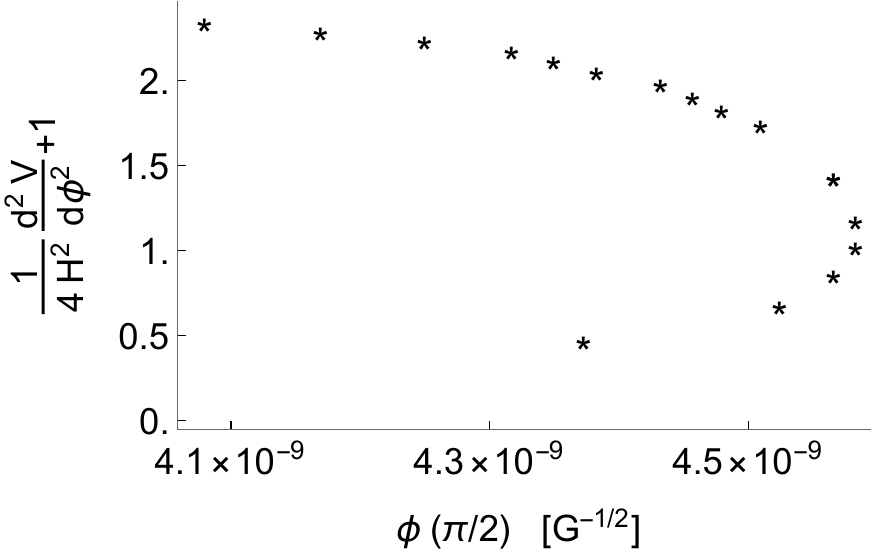}\end{minipage}}
    \caption{Top: Eq.\eqref{eq:leftbound} for a polynomial potential Eq.\eqref{eq:quadrpotential} for $b=\frac{1}{2\sqrt{6} }, c=1$ (black),$b=\frac{1}{40\sqrt{6}}, c=1$ (red), $b=-\frac{1}{2\sqrt{6}}, c=1$ (blue), $b=\frac{1}{2\sqrt{6}}, c=-1$ (green). Bottom: Eq.\eqref{eq:leftbound} for the Higgs potential.}
    \label{fig:quadrpi2}
\end{figure}

 To test our results, consider a scalar field theory with polynomial potential
\bea
\label{eq:quadrpotential}
V(\phi)=\dfrac{703}{100}\left(-\dfrac{\phi^2}{2}- b \phi^3+c\dfrac{\phi^4}{4}\right)
\eea
which has $\phi_{top}=0$ and $V''(\phi_{\rm top})=-7.03$. Eq.\eqref{eq:chap3:vds} is monotonic for $b\rightarrow 0$ and $4 H^2>7.03$. Moreover, for $b>0$, the monotonicity breaks on the right  of $\phi_{top}$ and thus one has $\phi(\pi/2)<\phi_{\rm top}$. Analogously, for $b<0$ one has $\phi(\pi/2)>\phi_{\rm top}$. The value of Eq.\eqref{eq:leftbound} as a function of $\phi(\pi/2)$ for various $b$, $c$ is reported in Fig.\ref{fig:quadrpi2}, on the left, and shows that:
\begin{itemize}
    \item positive quartic powers give an upper bound (blue, red and black lines), while negative ones a lower bound (green line)
    \item positive cubic powers give $\phi(\pi/2)>0$ (blue line), while negative ones give $\phi(\pi/2)<0$ (green, red and black lines).
    \item the smaller is the cubic power, the closer is $\phi(\pi/2)$ to $\phi_{\rm top}$.
\end{itemize}
\vspace{0.5cm}

As further example, consider the Higgs potential with two-loop corrections \cite{Gregory:2016xix}
\bea
\label{eq:chap1:higgspotential}
 V(\phi)=\dfrac{\lambda(\phi)}{4} \phi^4
 \eea
 with
  \begin{equation}\label{eq:chap1:runninglambda}\lambda(\phi)=\lambda^*+\gamma \ln\left(\phi\right)^2+\beta \ln (\phi)^4\end{equation} where $\phi$ is measured in Planck mass units $G=1$ and compute $V''(\phi)$ as a function of $\phi$. It vanishes at $\phi_{\rm fv}$, it gets positive in a small region with $\phi<\phi_{\rm top}$ and it is negative for $\phi\geq\phi_{\rm top}$, with
$V''(\phi)\rightarrow-\infty$ for $\phi\rightarrow +\infty$. Then Eq.\eqref{eq:leftboundbarrier}, as a function of $\phi$, is always positive around the false vacuum, and this region is larger the larger is $H$. Then, Eq.\eqref{eq:chap3:vds} is never monotonic. The quartic derivative instead is negative definite apart from a small region around the false vacuum of order $\phi\sim 10^{-9}\, G^{-1/2}$. The potential barrier is infinitely large, and so this may be regarded as a small monotonicity breaking for an otherwise negative definite $V''(\phi)$, for small enough $H$. Thus, $H$ has a lower bound which is approximately determined by Eq.\eqref{eq:leftbound}, i.e. $H^2_{\rm b}\sim1.5\times 10^{-20}\,G^{-1}$. The upper bound on $H$ can only be found numerically, as there is no true vacuum state (see Fig.\ref{fig:quadrpi2}). It is found to lie at $H^2 \sim 3.4\times 10^{-20}\,G^{-1}$.\\
 
 \section{Adding Einstein-Hilbert gravity}
 \label{sec:EHds}
 The bounds on $H$ in scalar field theories with Einstein-Hilbert gravity are similar to the ones found in the previous section. In this case the action is 
 \bea\label{eq:chap2:action}S_E=2 \pi^2 \int dt \,\rho(t)^3 \left[-\dfrac{M_P^2 R}{2}+\dfrac{1}{2} g_{\mu\nu}\nabla^{\mu}{\phi}\nabla^{\nu}{\phi}+V(\phi)\right]. \eea
 Here $M_P$ indicates the reduced Planck mass and $g_{\mu\nu}$ is the metric with line element Eq.\eqref{eq:line}. The Ricci scalar $R$ is given by
 \bea
 \label{eq:chap2:ricci}
 R=-6\,\dfrac{\dot{\rho}^2-1+\ddot{\rho}\rho}{\rho^2}
 \eea
and the Hubble parameter is defined as
 \bea
 \label{eq:chap2:hubble}
 H^2\equiv-\dfrac{\dot{\rho}^2-1}{\rho^2}.
 \eea
 The scalar field equation of motion is 
 \bea\label{eq:chap2:eom}\ddot{\phi}+\dfrac{3\dot{\rho}\,\dot{\phi}}{\rho}=V'\eea
 while the $tt-$component of the Einstein equations is
 \bea
 \dot{\rho}^2=1+ \dfrac{\rho^2}{3M_P^2} \left(\dfrac{\dot{\phi}^2}{2}-V(\phi)\right).\label{eq:chap2:eom2}\eea 
 $V(\phi_{\rm fv})>0$ in order to have a de Sitter false vacuum state. 
In the asymptotic region on the bounce (that is, for $Ht\sim \pi$) the scale factor is determined by Eq.\eqref{eq:scaleds}. 
To avoid any quantum gravity effects, all scales on the bounce should be smaller than the Planck mass. By Eq.\eqref{eq:chap2:eom2} one has then that
 \bea
 \label{eq:chap3:scaleds}
 \rho(t)\approx\dfrac{1}{\bar{H}}\sin(\bar{H} t)
 \eea
to lowest order at \emph{all} times on the bounce, with $\bar{H}$ determined by Eq.\eqref{eq:chap2:hubble} for $\phi=\bar{\phi}$ (as defined in Sect.\ref{sec:backgrounddS}), i.e.
\bea
\label{eq:boundaryH}
\bar{H}^2=\dfrac{4 V(\bar{\phi})}{3 M_P^2},\qquad V'(\bar{\phi})>0.
\eea 
Under this approximation, bounds on $H$ may be estimated analogously than in the case of a scalar field on a fixed de Sitter background, the only difference being that $\bar{H}$ is to be chosen among the ones satisfying Eq.\eqref{eq:boundaryH}.
If no $\bar{H}$ satisfies the bounds, than a bounce is excluded. Notice that this differs from  Eq.\eqref{eq:chap2:nec} with 
 \bea
H^2=\dfrac{V(\phi)}{3M_P^2}
\eea 
as having $V(\phi)=0$ in the potential barrier makes Eq.\eqref{eq:chap2:nec} unbounded from below. In the case of Higgs decay, one has $V(\phi_{\rm fv})=0$ and thus de Sitter space emerges only by adding a cosmological constant in the action. Thus, the value of $\bar{H}$ is not constrained by the scalar field potential, and so bounds on it are the same as in Sect.\ref{sec:backgrounddS}.
 \section{Modified gravity}
 In this Section, we generalize the bounds derived in Sect.s\ref{sec:backgrounddS}-\ref{sec:EHds} to modified gravity theories with a non minimal coupling $\xi \phi^2 R$ and a quadratic Ricci term $\alpha R^2$. The line element is again given by Eq.\eqref{eq:line}, and the Euclidean action is
 \bea\non
 \label{eq:actionscale}
     S_E&=&2\pi^{2}\int_0^{+\infty}dt\rho(t)^3\Bigg(\dfrac{\dot{\phi}^2}{2}+ V(\phi)-\dfrac{M_{\rm P}^2}{2} R-\dfrac{\xi}{2} \phi^2 R+\dfrac{\alpha}{36} R^2\Bigg)
 \eea 
 with Ricci scalar given by Eq.\eqref{eq:chap2:ricci}.
The scalar field equation of motion is 
\bea
 \label{eq:eom2scale}\ddot{\phi}+\dfrac{3\dot{\rho}\,\dot{\phi}}{\rho}=V'(\phi)-\xi \phi R.\eea
 while the $tt-$component of the Einstein equations is
 \bea
 \label{eq:eom1scale}\dot{\rho}^2=1+\rho^2 \, \dfrac{\dfrac{\dot{\phi}^2}{2}-V(\phi)+\dfrac{\alpha}{36} R^2+\left(\dfrac{\alpha}{3}\dot{R}-6 \xi\,\phi\,\dot{\phi}\right)\dfrac{\dot{\rho}}{\rho}}{3\left(M_{\rm P}^2+\xi \phi^2-\dfrac{\alpha}{9} R\right)}.\eea
 Taking the trace instead one gets
 \begin{gather}
    \alpha \left(\ddot{R}+\dfrac{3\dot{\rho}}{\rho}\dot{R}\right)=-3 M_{\rm P}^2R- 3 \xi (1+6 \xi)\phi^2R+3 \dot{\phi}^2 (1+6 \xi)+12 V(\phi)+18 \xi \phi V'(\phi)\notag\\\label{eq:tracescale}
 \end{gather}
 where derivatives of $R$ in $t$ indicate the presence of an additional scalar degree of freedom in the gravitational sector if $\alpha\neq 0$. In the following, we will switch on alternatively $\xi$, $\alpha$, in order to see their effect on the bounds on $H$. We start by considering a theory with a non minimal coupling and then turn to quadratic theories of gravity.
\subsection{Non-minimal coupling}
\label{sec:nonm}
As described in Sect.\ref{sec:EHds}, vacuum decay from de Sitter space requires that the right-handside of Eq.\eqref{eq:chap2:eom} and Eq.\eqref{eq:chap2:hubble} to be positive at $t=\pi \bar{H}^{-1}$. These conditions determine the possible values of $\bar \phi$ (and, thus, $\bar{H}$) one should consider when testing for the viability of vacuum decay. If $\xi\neq 0,\,\alpha=0$ these conditions read
\bea
\label{eq:Rnonm}
V'(\bar{\phi})-\xi \bar{\phi} \bar{R}>0\qquad \bar{R}=12 \bar{H}^2=\dfrac{12 V(\bar{\phi})+18 \xi \bar{\phi} V'(\bar{\phi})}{3 M_P^2+ 3 \xi (1+6 \xi)\bar{\phi}^2 }>0.
\eea
 If the theory has no Einstein-Hilbert term, also other conditions should be imposed. In particular, $\xi>0$ is required in order to have a positive definite effective squared Planck mass $\xi \phi^2$ and $\phi\neq 0$ on the bounce, so that the solution to Eq.\eqref{eq:eom1scale} is regular.\\
We expect that a non-minimal coupling also affects bounds on $H$, as it alters the scalar field equation of motion. Considering the same approximations adopted in Sect.\ref{sec:EHds} and using Eq.\eqref{eq:chap3:scaleds}, Eq.\eqref{eq:chap3:fpi2} one has that
\bea
\label{eq:boundnonm}
V''(\phi)+4(1-3 \xi) \bar{H}^2
\eea
replaces
\bea
V''(\phi)+4  \bar{H}^2
\eea
in the whole calculation. As a result, thus, $H_{\rm b}$ is also a function of $\xi$, and according to Eq.\eqref{eq:boundnonm}, it gets larger for increasing $\xi$. In order to determine when a bounce is allowed, one should consider all possible values of $\bar{H}$, and compare them to the bounds predicted by the discussion in Sect.\ref{sec:backgrounddS}, with Eq.\eqref{eq:leftboundbarrier} replaced by Eq.\eqref{eq:boundnonm}. Notice that, for $\xi\geq1/3$, the function
\bea
V'(\phi)+4 \bar{H}^2 (\phi-\phi_{\rm top})(1- 3 \xi)
\eea
is positive definite for $\phi<\phi_{\rm top}$ while negative for $\phi>\phi_{\rm top},$ independently of $\bar{H}$. As discussed in Sect.\ref{sec:backgrounddS}, this condition is sufficient to exclude the presence of bounces. Then, one has that no bounce exists for such values $\xi.$\\

\begin{figure}
    \centering
    \mbox{\begin{minipage}{0.5\textwidth}
    \includegraphics[scale=0.5]{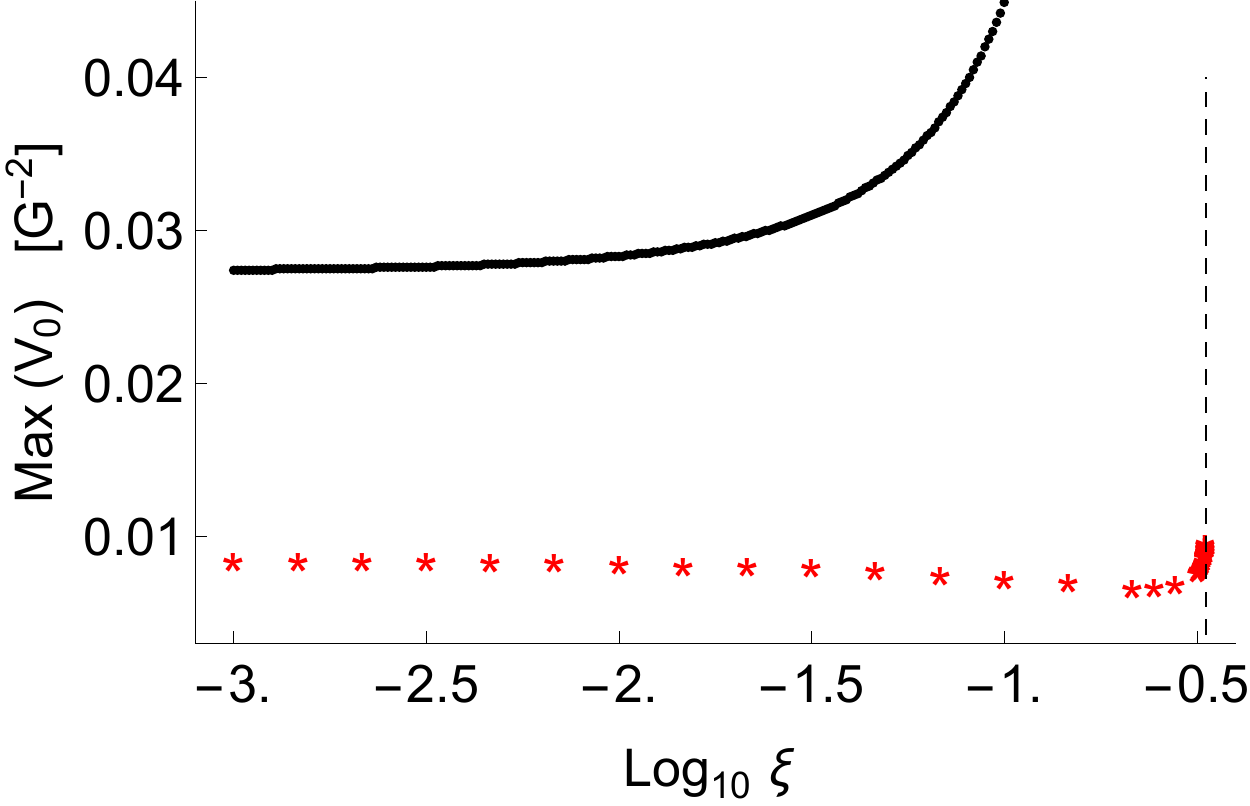}
    \end{minipage}
    \begin{minipage}{0.5\textwidth}
    \includegraphics[scale=0.5]{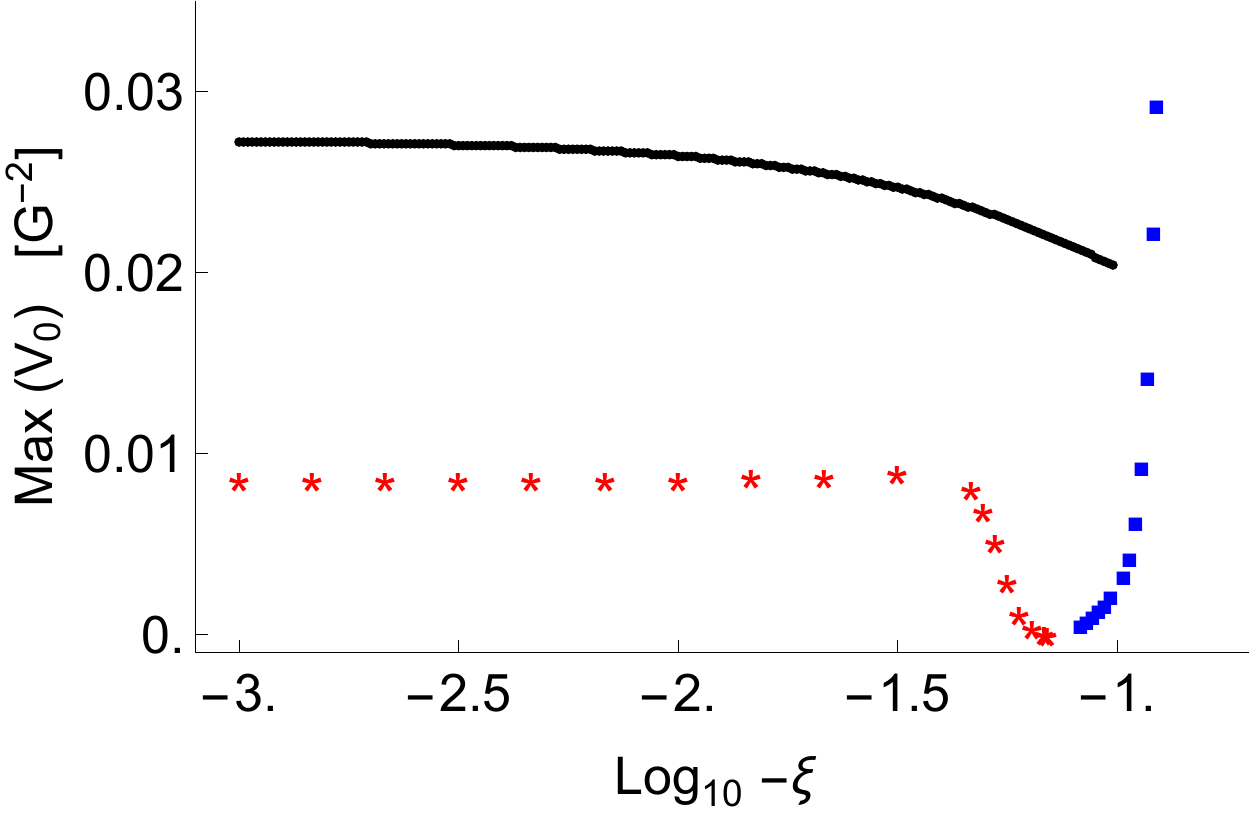}
    \end{minipage}}
    \caption{Theoretical (black line) and numerical (red stars, blue squares) bound on $V_0$, for $\xi>0$ (left) and $\xi<0$ (right). The potential is Eq.\eqref{eq:chap4:examplequart}.}
    \label{fig:boundnonm}
\end{figure}

\begin{figure}
    \centering
    \mbox{
    \begin{minipage}{0.5\textwidth}
    \includegraphics[scale=0.5]{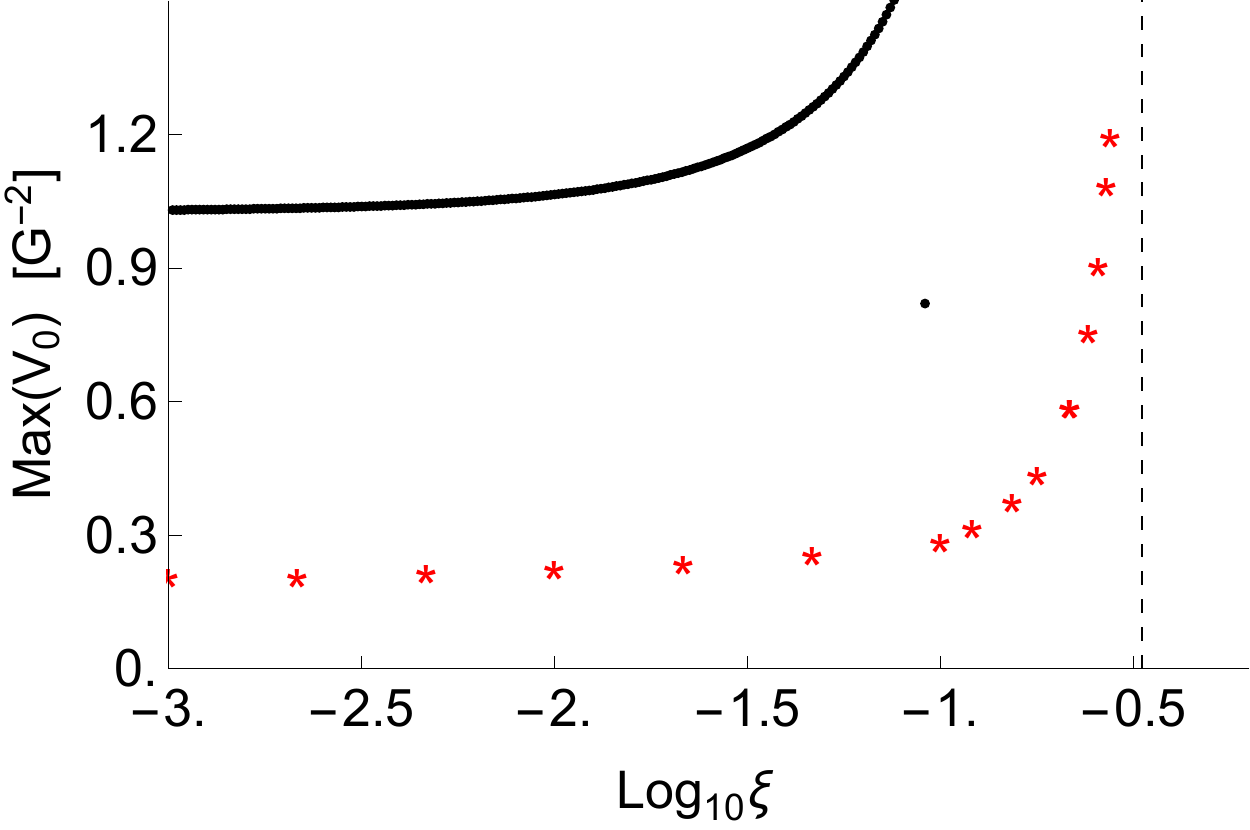}
    \end{minipage}\begin{minipage}{0.5\textwidth}
    \includegraphics[scale=0.5]{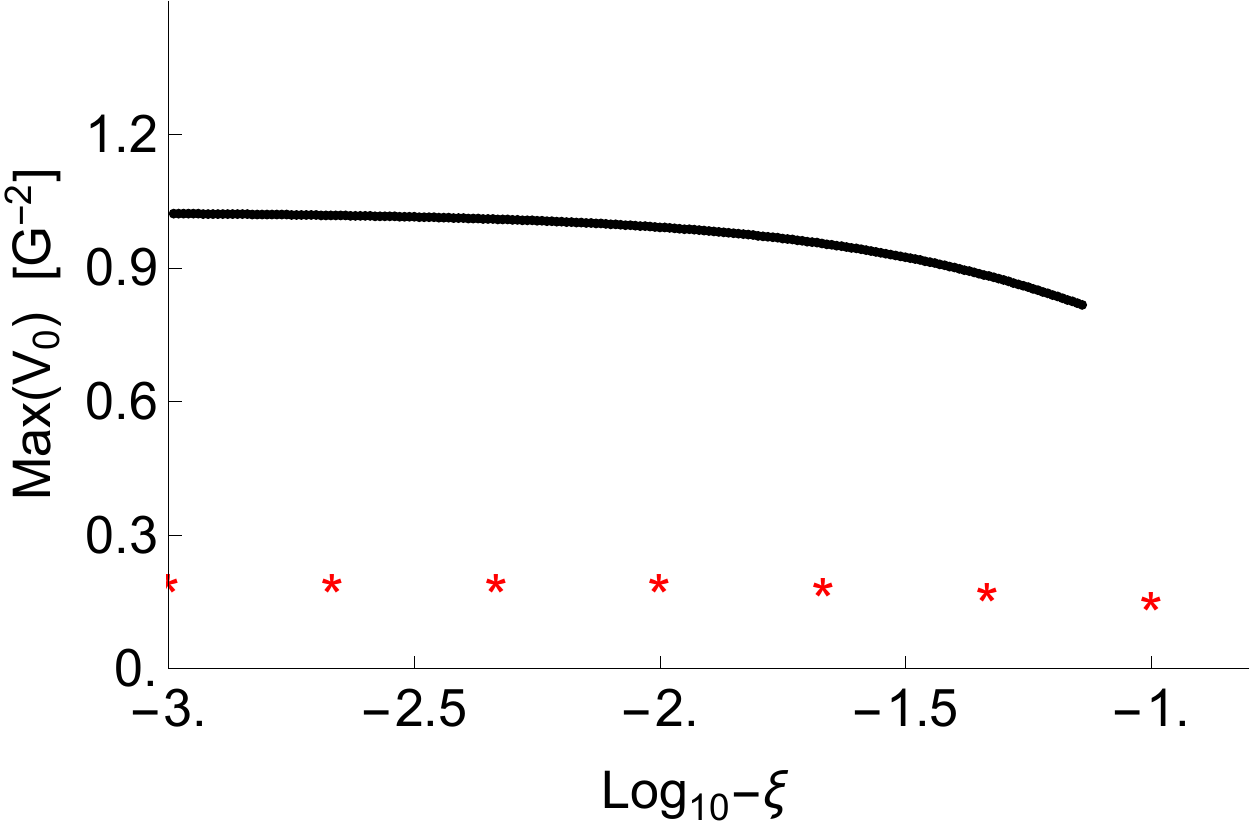}
    \end{minipage}}
    \caption{Theoretical (black line) and numerical (red stars) bound on $V_0$, for $\xi>0$ (left) and $\xi<0$ (right). The potential is Eq.\eqref{eq:chap4:examplequadr}.}
    \label{fig:boundnonmmass}
\end{figure}

\vspace{0.5cm}
The bounds on $\bar{H}$ just derived are compared with the numerical one in two toy models in Fig.\ref{fig:boundnonm} and Fig.\ref{fig:boundnonmmass} ($\xi>0$ on the left, $\xi<0$ on the right), with 
\bea
\label{eq:chap4:examplequart}
V(\phi)=V_0+\dfrac{703}{100}\phi^2\left(-\dfrac{\phi^2}{2}-\phi^3+\dfrac{\phi^4}{4}\right)
\eea
and \bea
\label{eq:chap4:examplequadr}
V(\phi)=V_0+\dfrac{703}{100}\left(-\dfrac{\phi^2}{2}-\dfrac{\phi^3}{2 \sqrt{6}}+\dfrac{\phi^4}{4}\right)
\eea respectively. Both potentials have $\phi_{\rm top}=0$ and $\phi_{\rm fv}<0$.
The former has
\bea
V''(\phi_{\rm top})=0,
\eea
and thus the upper bound on $\bar{H}$ is determined by the requirement that
 \bea
 \label{eq:examplequadr}
V''(\phi)+4(1-3\xi) \bar{H}^2<0
 \eea
 somewhere in the potential barrier. The latter has 
 \bea
 V''(\phi_{\rm top})\neq 0
 \eea
and the upper bound is set as 
\bea
\label{eq:boundbeyondtop}
V''(\phi_{\rm top})+4(1-3\xi)H_{\rm b}^2=0.
\eea
The red stars and blue squares represent the numerical upper bound, i.e. the value of $\bar{H}$ above which no bounce was found numerically for $\phi<0$ and $\phi>0$ respectively. The black line represents the theoretical bound on $\bar{H}$ for the quadratic potential, while the requirement
 \bea
V''(\phi)+4(1-3\xi)\bar{H}^2<0\qquad \text{for}\qquad \phi\in [\phi_{\rm fv},\phi_{\rm top}]
 \eea
in the quartic case, as considering the whole potential barrier set the bound at too high values of $V_0$ to be properly represented in the plot. The dashed line marks the value $\xi=1/3$. The expected behaviour is, roughly, that increasing $\xi$ softens the bound on $\bar{H}$, and this is indeed observed in Fig.\ref{fig:boundnonm}-\ref{fig:boundnonmmass}. $\bar{H}$ is driven to $+\infty$ as $\xi=1/3$ is approached in both cases. If $\xi$ is negative, the bounce disappears for $\xi<-0.1$ in the quadratic case, as, for large enough $|\xi|$ and $\xi<0$, the first of Eq.\eqref{eq:Rnonm} is always violated for $\phi<\phi_{\rm top}$. In the quartic case instead $\bar{\phi}$ migrates to the region $\phi>\phi_{\rm top}$, in which the first of Eq.\eqref{eq:Rnonm} is satisfied for $\bar{R}$ such that
\bea
\lim_{\phi\rightarrow 0}\dfrac{V(\phi)-V_0}{\xi \bar{R}}<1.
\eea
In this case, thus, the bounce survives for larger values of $\xi$. \\

\vspace{0.5cm}
As further example consider the Higgs field with an Einstein-Hilbert term, a cosmological constant $V(\phi_{\rm fv})$ and a non-minimal coupling. The first of Eq.\eqref{eq:Rnonm} is satisfied for some $\phi$ and inflationary values of $\bar{H}$, ($\bar{H}>10^{-6}$ in units $G=1$) if $\xi< 10^{-9}$. If $\xi>0$, the non-minimal coupling leaves $H_{\rm b}$ almost untouched, as, for such $\bar{H}$ ($\bar{H}\sim 10^{-10}$), the non-minimal coupling is very small with respect to other scales in the system. Nonetheless, the upper bound may only be predicted numerically, as there is no true vacuum state, and thus a bounce cannot be in principle excluded. For $\xi<0$, instead, the region available for $\bar{\phi}$ grows as $\xi$ gets smaller, and thus a bounce is in principle allowed for all values of $\xi$, with $H_{\rm b}$ getting smaller as $\xi$ decreases. $H_{\rm b}$ would cross the current cosmological constant value at extremely large values of $\left|\xi\right|$ ($\xi\sim -10^{100}$) for which $V'(\phi)$ is positive definite for $\phi\leq M_P$. Thus Higgs decay with non-minimal coupling is excluded in the current Universe and in the semi-classical approximation.\\

 \begin{tikzpicture}[overlay,remember picture]
 \node at (4.8,10.2){\begin{tikzpicture}\draw  [color=white,fill=white](0,1)--(1,1)--(1,0)--(0,0)--(0,1);\end{tikzpicture}};
 \end{tikzpicture}
\begin{figure}
    \centering\mbox{\begin{minipage}{0.5\textwidth}
    \includegraphics[scale=0.5]{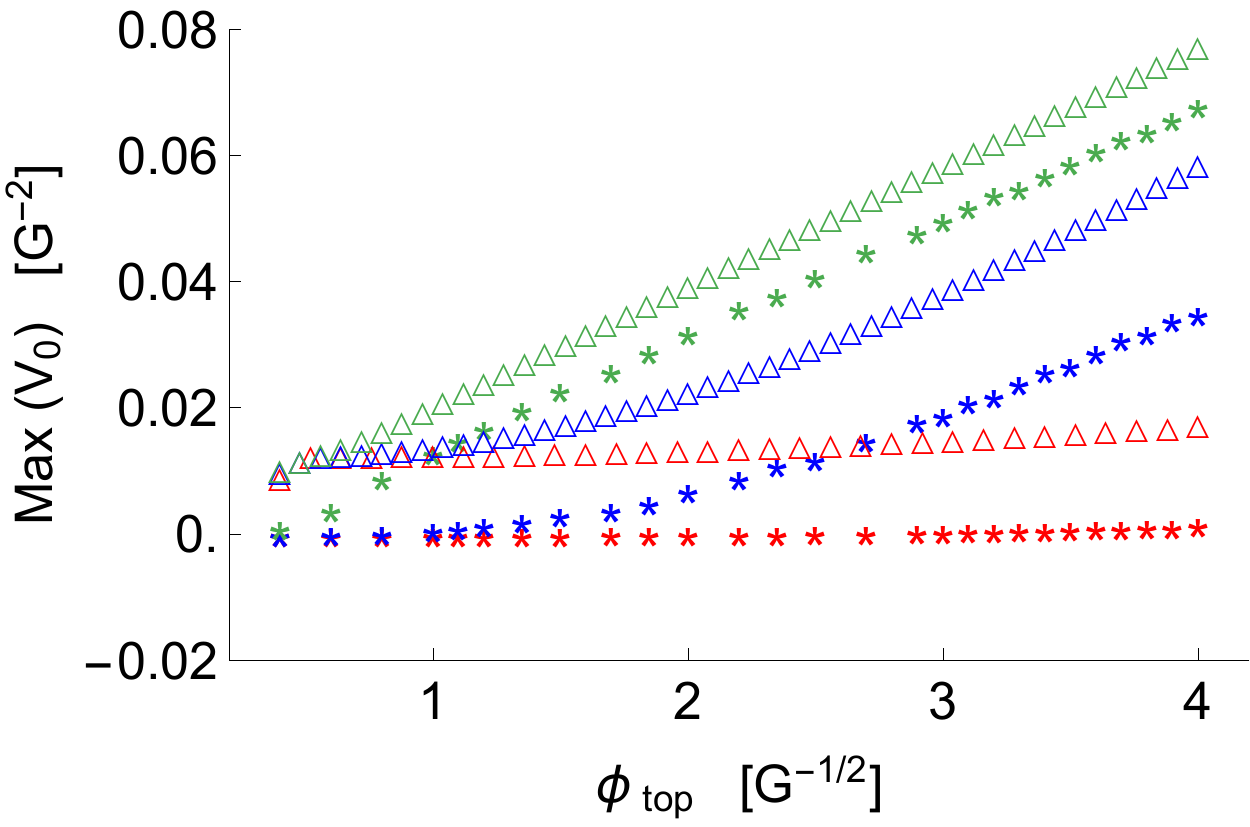}
    \end{minipage}
    \begin{minipage}{0.5\textwidth}
    \includegraphics[scale=0.5]{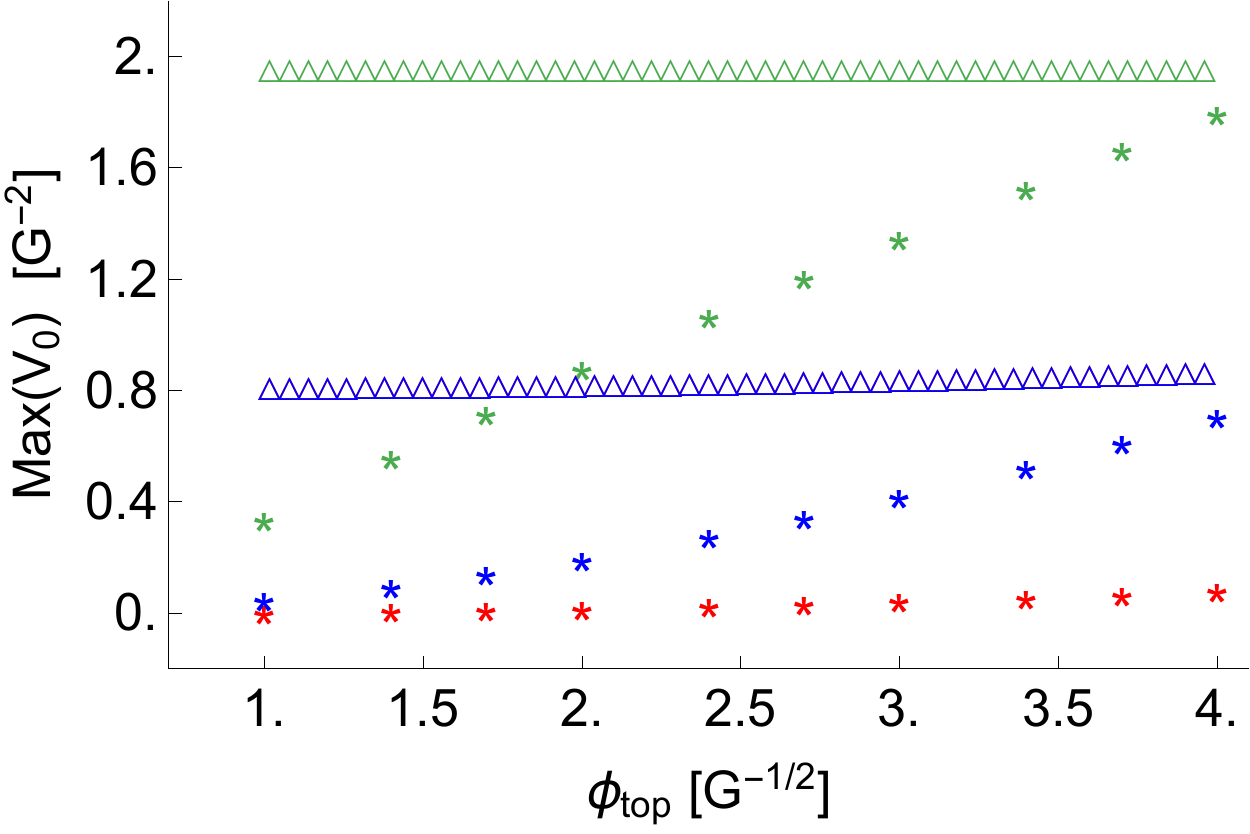}
    \end{minipage}}
    \caption{Left: Bound on $V_0$ as a function of $\phi_{\rm top}$ for $\log \xi =-3,-2,-1$ (red, blue and green respectively). The potential is Eq.\eqref{eq:chap4:examplequart} with shifted false vacuum value $\phi\rightarrow \phi-\phi_{\rm fv}$ so that $\phi_{\rm fv}>0.$ Stars represent the numerical bound, which is determined by the highest value of $V_0$ for which a bounce was detected numerically. Triangles represent the bound Eq.\eqref{eq:boundbeyondtop}. Right: same as on the left, in a theory with potential Eq.\eqref{eq:examplequadr} with shifted false vacuum value $\phi\rightarrow \phi-\phi_{\rm fv}$ so that $\phi_{\rm fv}>0.$ Triangles represent the theoretical bound Eq.\eqref{eq:boundbeyondtop}.}
    \label{fig:boundnonmsi}
    
\end{figure}
\vspace{0.5cm} 

Consider now briefly the limit $M_P\rightarrow0$. As stated above, one needs $\xi>0$ and $\phi\neq0$ on the bounce, besides Eq.s\eqref{eq:Rnonm}. Moreover, as the bounds described in Sect.\ref{sec:backgrounddS} are independent of $M_P$, similar results to the ones just described in the $M_P\neq 0$ case should apply (see Fig.\ref{fig:boundnonmsi} for an example).
\subsection{Quadratic gravity}
\label{sec:chap4:quadr}
\begin{figure}
    \centering
    \mbox{
    \begin{minipage}{0.5\textwidth}
    \includegraphics[scale=0.45]{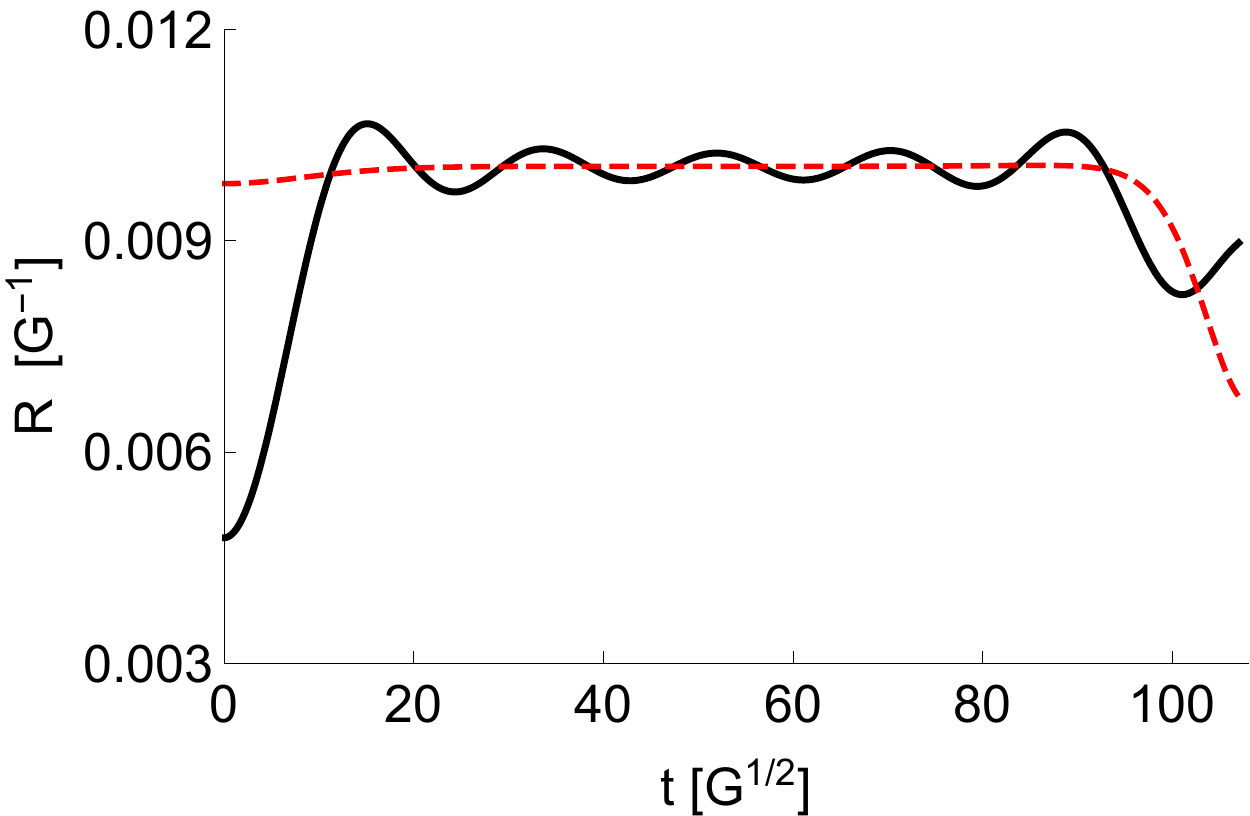}
    \end{minipage}
    \begin{minipage}{0.5\textwidth}
    \includegraphics[scale=0.5]{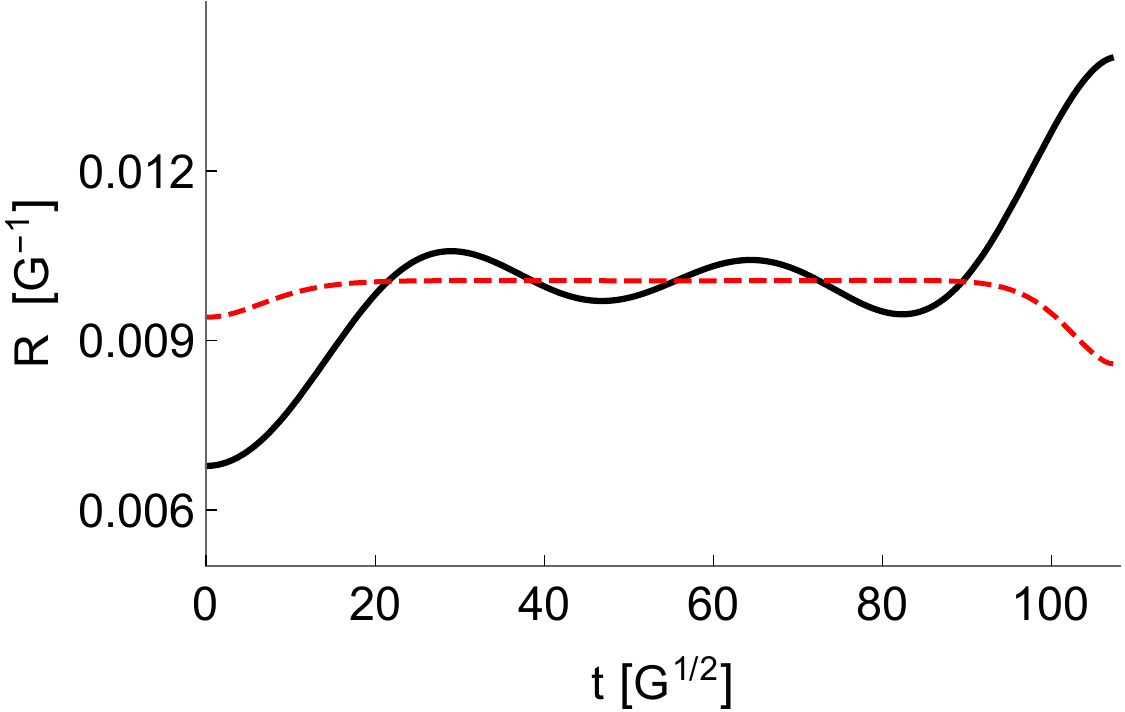}
    \end{minipage}}\\
    \includegraphics[scale=0.45]{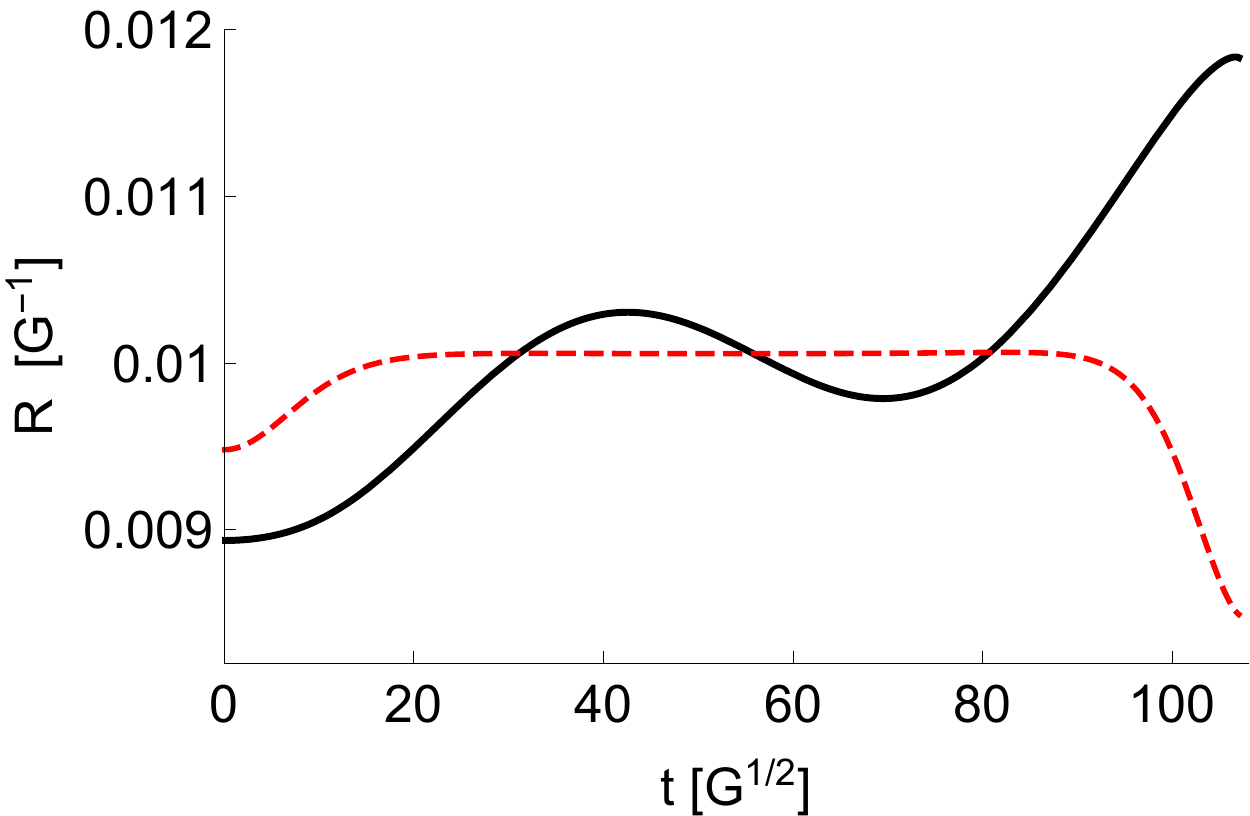}
    \label{fig:quadr}\caption{Ricci scalar trajectory on the bounce for a scalar field theory with potential Eq.\eqref{eq:chap4:examplequart} with an Einstein-Hilbert term and a quadratic Ricci scalar for $\alpha=1, 3.6,8$ respectively on the top left, right and on the bottom. One can notice that $R$ is closed to the fixed point for $\bar{H}t\geq \pi/2$, and that it slightly deviates when $t\approx \pi \bar{H}^{-1}$ is approached.}
\end{figure}

Quadratic gravitational terms have important consequences on bounds on the Hubble constant that are mainly related to the Ricci scalar being a propagating degree of freedom. First of all, $R$ needs to satisfy similar boundary conditions as the scalar field, namely
\bea
\dot{R}(0)=0\qquad \dot{R}(\pi \bar{H}^{-1})=0.
\eea 
Notice also that, in this case, $\bar{R}$ is not fully determined by $\bar{\phi}$, and that the scalar field equation of motion depends on $\rho(t)$ only through the friction term which is, to lowest order, determined by Eq.\eqref{eq:chap3:scaleds}. Thus, a bound on $\bar{H}$ should be computed as in Sect.\ref{sec:backgrounddS} and compared to \emph{all} real values of (constant) $\bar{H}$.\\ In principle, one may also try to impose a similar bound than the ones described in Sect.\ref{sec:backgrounddS} on $\bar{R}$, setting Eq.\eqref{eq:chap3:scaleds} in the friction term of Eq.\eqref{eq:tracescale}, and treating it as any scalar degree of freedom. The equations of motion though are entangled through the trace of the stress-energy tensor, which makes a clean analysis of an upper bound for $\bar{H}$ complicated. Notice though that the curvature $-3 M_P^2\alpha^{-1}$ decreases for increasing $\alpha$, and thus we expect that the bounce disappears for sufficiently large $\alpha$, supposing constant $\bar{H}$ (whose value is for example set by the physics of the false vacuum state) and $V(\bar{\phi})$.\\
In principle, one has a bounce for de Sitter false vacua, in contrast to flat space ones (see \cite{nostro1,nostro2} for details). Nonetheless, the smaller is the cosmological constant, the more difficult is to make predictions on the on-shell action, as $R$ oscillates on the bounce as long as spacetime is flat to a good approximation, i.e. for $t\ll \pi (2 \bar{H})^{-1}.$ This makes a calculation of the tunneling exponent for Higgs decay in the current universe particularly difficult, even numerically. Moreover, it is questionable whether such bounce would contribute to the false vacuum decay, due to the high number of oscillations in $R$, for the same reasons concerning the oscillating instantons described in the Introduction.\\
Anyway, if Eq.\eqref{eq:chap3:scaleds} holds, the Ricci scalar should be sufficiently constant before approaching $t=\pi \bar{H}^{-1}$. This implies that it should stay close to a fixed point of Eq.\eqref{eq:tracescale}. On the other hand, as stressed in Sect.\ref{sec:backgrounddS}, propagating scalar degrees of freedom do not reach fixed points of the equations of motion at the boundary. So, in order to have a bounce, one should additionally require that the velocity of the Ricci scalar is small
\bea
\label{eq:boundquadr}
\dfrac{\lvert\bar{H}\dot{R}\rvert}{\bar{R}^2}\ll 1.
\eea
$\bar{R}$ is determined as (see Eq.\eqref{eq:eom1scale}) 
\bea
\label{eq:constR}
-\dfrac{\bar{R}}{12}=\dfrac{-36V(\bar{\phi})+\alpha \bar{R}^2}{12(9M_P^2-\alpha \bar{R})}
\eea
where $\bar{ H}\dot{R}$ was neglected with respect to $\bar{R}^2$ and $\dot{\phi}^2\ll V(\phi)$ sufficiently close to the boundary. Eq.\eqref{eq:constR} gives
\bea
\bar{R}=\dfrac{4 V(\bar{\phi})}{M_P^2}
\eea
which is also the fixed point of Eq.\eqref{eq:tracescale}. Estimating the magnitude of $\dot{R}$ and $\ddot{R}$ from Eq.\eqref{eq:tracescale} one has that small deviations from the fixed point  give small velocities, namely
\bea
\alpha \bar{H} \dot{R}\sim 4 A V(\bar{\phi})\qquad \text{for}\qquad \bar{R}\sim 4 A V(\bar{\phi}) M_P^{-2}
\eea
where $A\lesssim O(1)$. Then
\bea
\dfrac{\bar{H}\dot{R}}{\bar{R}^2} = \dfrac{A M_p^4}{4 \alpha V(\bar{\phi})}
\eea
and thus it depends, as expected, on the magnitude of deviations from the fixed point. By setting $M_P^2=0$ instead Eq.\eqref{eq:constR} turns into the requirement $V(\bar{\phi})=0$, contradicting that, in general, $V(\bar{\phi})>0$. Thus a bounce does not exist if $\alpha\neq 0,\,\xi=0$ and $M_P=0$. In Fig.\ref{fig:quadr} the bounce for $R$ (black line) in a theory with potential Eq.\eqref{eq:chap4:examplequart} is reported as a function of $t$ for $\alpha=1,3.6$ and $8$. This is compared to the fixed point (red line)
\bea
R=\dfrac{4V(\phi)-\dot{\phi}^2}{M_P^2}.
\eea
Near the boundary
\bea
\dfrac{\lvert\bar{H}\dot{R}\rvert}{\bar{R}^2}\sim 0.1, 0.06,0.02\ll1
\eea
respectively. No bounce with $M_P^2=0$ was found.
\\
\subsection{Non-minimal coupling and quadratic Ricci scalar}
\label{sec:chap4:quadrnonm}
As described in Sect.\ref{sec:nonm}, the non-minimal coupling constrains the bounce by shifting the bound on $\bar{H}$ according to Eq.\eqref{eq:boundnonm}. When quadratic terms are added instead, the Ricci scalar velocity should be very small near the boundary. If both $\xi\neq0$ and $\alpha \neq 0$, the bounds on $\bar{H}$ carry over directly from the $\alpha=0$ case, apart from the fact that $\bar{R}$ is a free parameter, and not completely fixed by the scalar field. The bound on $\dot{R}$ instead may be significantly modified, especially in the $M_P=0$ limit, in which there is no bounce if $\xi =0$. Imposing Eq.\eqref{eq:constR} in this case gives
\bea
\bar{R}=\dfrac{4 V(\bar{\phi})}{M_P^2+\xi \bar{\phi}^2}.
\eea
Plugging this result in non-derivative terms in $R$ in Eq.\eqref{eq:tracescale} and using it to estimate $\bar{H}\dot{R}$ one gets
\bea
\label{eq:approxdotR}
\lvert\bar{H}\dot{R}\rvert\approx \left\lvert\dfrac{6\xi\bar{\phi}}{\alpha}  \left(V'(\bar{\phi})-\dfrac{4 \xi \bar{\phi} V(\bar{\phi})}{M_P^2+\xi \bar{\phi}^2}\right)\right\rvert
\eea
setting both terms on the right-hand side to be separately small according to Eq.\eqref{eq:boundquadr} implies
\bea
\label{eq:boundxi}
\left\lvert\dfrac{\xi}{\alpha}\right\rvert\ll \left\lvert\dfrac{8V(\bar{\phi})^2}{3\bar{\phi}V'(\bar{\phi})(M_P^2+\xi \bar{\phi}^2)^2}\right\rvert\qquad \left\lvert\xi \bar{\phi}^2\right\rvert\ll\dfrac{2\alpha V(\bar{\phi})}{3(M_P^2+\xi \bar{\phi}^2)}
\eea

If $M_P=0$ instead
\bea
\bar{R}=\dfrac{4 V(\bar{\phi})}{\xi \bar{\phi}^2}.
\eea
In the semi-classical approximation one needs $\xi \phi_{\rm fv}^2 M_P^2\gg V(\bar{\phi})$, and thus Eq.\eqref{eq:boundquadr} gives
\bea
\label{eq:boundximp0}
\left\lvert\dfrac{\xi}{\alpha}\right\rvert\ll \dfrac{V(\bar{\phi})}{\xi^2 \bar{\phi}^4}\ll 1\qquad \text{and} \qquad \left\lvert \dfrac{\xi}{\alpha}\right\rvert\ll \left\lvert\dfrac{V(\bar{\phi})^2}{\xi^2 \bar{\phi}^5 V'(\bar{\phi})}\right\rvert
\eea

The bound Eq.\eqref{eq:boundxi}  was computed for a scalar field theory with potential Eq.\eqref{eq:chap4:examplequart} and $\alpha=1$. Using Eq.\eqref{eq:boundquadr} one finds $\lvert\xi \rvert\ll 0.1$, while a bounce was found numerically up to $\xi\sim 0.02$, for positive $\xi$. Moreover,
\bea
\lvert\bar{H}\dot{R}\rvert \sim 0.15 \bar{R}^2\eea
and $\lvert\bar{H}\dot{R}\rvert\sim 1.3\times 10^{-5}.$ Using Eq.\eqref{eq:approxdotR} one gets 
\bea
\lvert\bar{H}\dot{R}\rvert\sim 2 \times 10^{-5}
\eea
agreeing within a multiplication constants of $O(1)$ with the numerical calculation. No bounce instead was found in the $M_P=0$ limit.\\

 \section{Conclusions}
 \label{conclusions}
 Vacuum decay in de Sitter space is a process of great physical interest, as it allows ruling out cosmological models in the early and current Universe. It is important to keep the underlying field theory as general as possible, as beyond-the-Standard-Model and early universe physics are still unknown. In the present paper, we derived new bounds for this process, that allow ruling out vacuum decay for both sufficiently small and sufficiently large values of the Hubble parameter. While in previous work \cite{nostro1,nostro2} we adopted a boundary analysis to derive existence conditions for the bounce with flat space false vacua, this is not possible in the present case, as de Sitter space with Euclidean signature is compact. Then, one needs some information not only close to the boundary, but also about the scalar field evolution in between. We found that, in particular, the midpoint $\phi(\pi (2 H)^{-1})$ serves the purpose, as its acceleration corresponds to Eq.\eqref{eq:dotfpi2}, which resembles the existence conditions previously found in the literature. First, we focussed on vacuum decay of scalar fields on a de Sitter background and derive existence conditions for Coleman-de Luccia (i.e. non-oscillating) instantons. Our results recover the ones previously presented in the literature \cite{Balek:2003uu} and also partially explain numerical predictions \cite{Hackworth:2004xb}. When applied to the Higgs field, we find that a lower bound is predicted which excludes vacuum decay in the present universe. Analogous findings may be derived in theories with a scalar field and Einstein-Hilbert gravity, as long as no quantum gravity effects are involved, i.e. all scales are much smaller than the Planck mass. Adding a non-minimal coupling one finds that similar bounds may be found, which now depend also on the non-minimal coupling $\xi$. In particular, we find that there is no bounce for $\xi>1/3,$ which we observed in two numerical examples. Adding a quadratic gravity term does not affect bounds on $H$, apart from the fact that now there is an additional gravitational scalar degree of freedom, and thus it is not completely set by the scalar field. Nonetheless, we can impose additional conditions as, in order to avoid quantum gravity effects, the velocity of the Ricci scalar should be sufficiently small. This requirement allows to rule out decay for theories with $\alpha\neq 0,\,,\xi=0$ and $M_P=0,$ and to constrain it under conditions Eq.\eqref{eq:boundxi}-\eqref{eq:boundximp0} for theories with $\alpha \neq 0,\,\xi\neq 0.$

\subsection*{Acknowledgments}
This work was performed during S.\,V.\, doctoral studies, which were financially supported by the Italian National Institute for Nuclear Physics (INFN). This work has been partially performed using the software Mathematica.

\end{document}